\documentclass[a4paper,fleqn]{cas-dc}

\usepackage[numbers]{natbib}
\usepackage{graphicx}
\usepackage{caption}
\usepackage{subcaption}

\usepackage{gensymb}
\usepackage{hyperref}
\usepackage{amsmath}
\usepackage{array}
\usepackage{enumitem}
\usepackage{lipsum}
\setlist[itemize]{leftmargin=*}

\def\tsc#1{\csdef{#1}{\textsc{\lowercase{#1}}\xspace}}
\tsc{WGM}
\tsc{QE}
\tsc{EP}
\tsc{PMS}
\tsc{BEC}
\tsc{DE}

\begin{document}
\let\WriteBookmarks\relax
\def\floatpagepagefraction{1}
\def\textpagefraction{.001}

\shorttitle{Preprint}

\shortauthors{W Yang et~al.}

\title [mode = title]{A generalised multi-factor deep learning electricity load forecasting model for wildfire-prone areas}

\author[1]{Weijia Yang}[type=editor,
                        orcid=0000-0002-3802-7722]         

\cormark[1]


\ead{weijia.yang@eng.ox.ac.uk}

\ead[url]{https://eng.ox.ac.uk/people/weijia-yang/}

\credit{Methodology, Software, Original draft writing, Review and editing}

\address[a]{University of Oxford, Department of Engineering Science, Oxford, OX1 3PJ, The United Kingdom}

\author[1]{Sarah N. Sparrow}

\credit{Review and editing, Methodology, Supervision}

\author[1]{David C.H. Wallom}

\credit{Conceptualization, Review and editing, Supervision}

\cortext[cor1]{Corresponding author}

\begin{abstract}
This paper proposes a generalised and robust multi-factor Gated Recurrent Unit (GRU) based Deep Learning (DL) model to forecast electricity load in distribution networks during wildfire seasons. The flexible modelling methods consider data input structure, calendar effects and correlation-based leading temperature conditions. Compared to the regular use of instantaneous temperature, the Mean Absolute Percentage Error (MAPE) is decreased by 30.73\% by using the proposed input feature selection and leading temperature relationships. Our model is generalised and applied to eight real distribution networks in Victoria, Australia, during the wildfire seasons of 2015-2020. We demonstrate that the GRU-based model consistently outperforms another DL model, Long Short-Term Memory (LSTM), at every step, giving average improvements in Mean Squared Error (MSE) and MAPE of 10.06\% and 12.86\%, respectively. The sensitivity to large-scale climate variability in training data sets, e.g. El Niño or La Niña years, is considered to understand the possible consequences for load forecasting performance stability, showing minimal impact. Other factors such as regional poverty rate and large-scale off-peak electricity use are potential factors to further improve forecast performance. The proposed method achieves an average forecast MAPE of around 3\%, giving a potential annual energy saving of AU\$80.46 million for the state of Victoria.
\end{abstract}



\newcommand{\SubItem}[1]{
    {\setlength\itemindent{15pt} \item[-] #1}
}

\maketitle

\section{Introduction}
To achieve the goal as set out in the Paris Agreement of minimising global mean temperature increase to under 2\degree C and ideally not more than 1.5 \degree C goal, a growing coalition of countries, businesses, and institutions have committed to implementing a low carbon energy transition thereby contributing to a net-zero emission future \cite{UNEP2022}. In the era of the energy transition, the diversification of the generation mix and electrification of household demand both raise great uncertainty in the power system. Meanwhile, extreme weather events are becoming more frequent and severe due to Climate Change. Climate-related disasters increased drastically from 3,656 events during the last 20 years of the $20^{th}$ century to 6,681 events in the first 20 years of the $21^{st}$ century \cite{E3602020}. The doubled risks of floods, 40\% more risk of storms, and a higher risk of wildfires, droughts, and heatwaves result in significant economic and social losses. The increasing climate variability also challenges the reliability and resilience of the energy system operation, e.g., more and more unintentional power blackouts and emergent power constraints are triggered due to extreme weather events. Between them, the emerging energy transition and the extreme weather events simultaneously bring uncertainty to the energy system, generating significant extra demands for power system operators.\\
In recent years, power grid damage attributed to wildfire risks has become more and more common, such as decreased line capacity, conductor line sag, unintentional power cuts, and a number of less apparent but no less impactful effects. In 2020, wildfires in South-East Australia burnt several million hectares of land in Victoria State, with over 20,000 households also being disconnected from the power system in New South Wales \cite{Yang2022}. Conversely, wildfires can be ignited by power system faults. According to a wildfire study in Victoria, Australia, electrically-caused wildfires are one of the six major causes, burning larger areas on average than other ignition sources \cite{Miller2017}. As Victoria, Australia, experiences annual extreme wildfire seasons, and the historical load data sets are accessible from public sources, we have selected this location as the case study region in our paper. The load on the Distribution Network (DN) in Victoria is modelled, assessed, and forecast by developing a multi-factor Deep Learning (DL) model aiming to improve the power system's reliability and security during extreme weather events.\\
To address the uncertainty in the power dispatch, academia and power operators have been exploring and applying different energy forecasting methods to dispatch power in advance and enhance the power system reliability under extreme event conditions. The most widely used load forecast methods in the real power system operation are conventional statistical models, e.g., linear regression-based models are used in the UK National Grid \cite{NG_forecast2017} and the PJM covered regions in the US \cite{PJM2023}. Although more and more factors are considered variables in the regression model, the uncertainty attributed to climate events and multi-source electrification is hard to capture using statistical models. Thus,  data-driven methods like Neural Networks (NN) have become an emerging field for solving load forecasting problems in recent research. Compared to traditional statistical methods, the NN-based model has the advantage of detecting the implicit and nonlinear relationship via the artificial neuron connection environment and can deal with high-dimension input issues to solve multi-factor problems \cite{Xiao2023}. Thus NN-based techniques, such as DL, are worthy of study to address uncertainty, multi-factor and nonlinear issues in load forecast for realistic and extreme use cases.

\subsection{Existing research in electricity load forecasting}
Existing research on electricity system load forecasting  can be categorised by different taxonomies. Existing studies may be classified by forecast period lengths, grid application levels, and methodologies. In terms of the forecast period, load forecasting can be divided into very-short-term load forecasting (VSTLF), short-term load forecasting (STLF), medium-term load forecasting (MTLF) and long-term load forecasting (LTLF) \cite{Eskandari2020}. Load predictions from a few minutes to an hour ahead and from an hour to a week ahead are defined as VSTLF and STLF, respectively, commonly used for the real-time electricity market balancing mechanism. For example, the half-hourly ahead demand forecast is used to determine which generators should be online to balance the regional demand as well as realise the most cost-effective system operation in the GB electricity markets \cite{ESO2023}. Load predictions with periods from weeks to a year ahead and several years ahead are classified as MTLF and LTLF, which can help with the generator unit maintenance scheduling and system expansion planning \cite{4554335}.\\
In the classification at grid level, the load forecast can be conducted on the country-wide, transmission network level, distribution network level and end-user level. At the country-level load forecast, LTLFs like the annual energy consumption and peak power are commonly regarded as the forecast goal. In \cite{Raheem2022}, the G20 country yearly energy consumption by sources was forecast using an adjacent accumulation grey model. The peak load of Korea was predicted using comparative methods in \cite{LEE2022122366}. At the transmission network level, some MTLFs are conducted, contributing to the transmission network planning. In \cite{9352301}, the Cambodian monthly peak load prediction for a year ahead was generated by linear regression and NN methods separately. At the DN level,  short-to-medium range load forecasts are studied. For example, the Portuguese medium voltage DN load for 24 hours to a week ahead was assessed using both Artificial Neural Networks (ANN) and a regression model in \cite{Chemetova2016}. On the household level, the STLF is usually analysed using the high temporal resolution data collected from smart meters. For instance, a household's energy use over the next few hours was forecast using DL approaches, analysing one-minute sampling rate data sets in \cite{8470406}.\\
In recent years, various methods have been developed to forecast energy consumption, including traditional time series statistical methods and Artificial Intelligence (AI) methods. Commonly used statistical load forecast techniques include Multiple Linear Regression (MLR), Autoregressive Integrated Moving Average (ARIMA), Semi-parametric Additive Models (SAM), and Exponential Smoothing Models (ESM) \cite{HONG2016914}. Support Vector Machines (SVMs) and ANN are AI techniques most widely studied in the load forecast research area. In \cite{6140142}, the daily peak load of Istanbul European Side was forecast using SVM, considering the past seven-day load, average daily temperature and electricity price as the model input. From the beginning of the $21^{st}$ century, some ANN-based basic forecast models have been designed and improved to capture the complex non-linear behaviour of energy use \cite{916876}, \cite{CHAN2006409}. In the past decade, ANN techniques have been further developed in two main directions: Recurrent Neural Network (RNN), with a specialisation in handling sequential data, and Convolution Neural network (CNN), which has the strength to capture the spatial features of data (image recognition) \cite{Xiao2023}. In \cite{8470406}, the basic RNN model and its derivative techniques: Long Short-Term Memory (LSTM) and Gated Recurrent Unit (GRU) models were studied to forecast household load in the next few hours. As the LSTM and GRU are techniques good at understanding the sequential data, LSTM and GRU are selected as the main model structure for the research reported in this paper, and the memory cell structures are explained in detail in the \nameref{method}.

\subsection{Research gap and main contribution}
Our paper proposes a multi-factor DL-based load forecast model via a case study of 8 real DNs in Victoria, Australia, achieving the average MAPE of around 3\% during wildfire seasons. In this subsection, each paragraph discusses one research gap and how we address the issue, reviewing the main contributions of our paper.
\begin{itemize}
\item So far, studies have yet to be implemented to assess the impact of adjusting input structures for the RNN-based load forecast model. In \cite{8470406} and \cite{YU2022120089}, the 24-hour ahead load data was default selected as the input for the DL forecast models since it is thought to be more appropriate in learning the characteristics exhibiting daily periodicity. However, the performance accuracy may already reach a stable level even with the load data input shorter than 24 hours. If we can find the shortest length of input that can achieve acceptable accuracy, the forecast model can be equipped with a higher computational efficiency.\\
- The input matrix structure of the multi-factor LSTM-based and GRU-based forecast model is adjusted to find the appropriate input data length with both good accuracy and less computational burden.

\item Based on the weekly periodicity in load, calendar effects have been applied to some load forecasts. However, little research has compared the different ways of using calendar labels and their impact on load forecasts. \cite{9701025} considered working days and non-working days as non-time series characteristics to assist in a Time Convolutional Network (TCN) and GRU combined model to forecast load for an industrial building. \cite{Zeng2018} generated six binary features and a zero feature to encode the calendar index, representing Monday to Sunday, respectively.\\
- Different ways of applying calendar labels on load forecast are compared to maximise the forecast accuracy advancement attributed to calendar label employment.

\item Climate factors have been added to several load forecast models, such as temperature, humidity, wind speed, precipitation, etc. In \cite{PENG2021111211}, the daily high \& low temperatures and the daily high \& low dew temperatures were set as the input features to help forecast the daily electricity consumption of a residential building in Shanghai, China. \cite{Zeng2018} used Pearson Correlation Coefficient to assess the relationship between climate variables and load, indicating that the maximum, minimum, and average temperature, air pressure and water vapour pressure are strongly correlated to load in a subtropical monsoon climate region. As one of the most significant climate variables in load forecast, most existing studies directly use the daily peak, mean, or instantaneous temperature \cite{6140142}, \cite{PENG2021111211}, \cite{MALDONADO2019105616}, \cite{Wang2021IEEE}. However, the relationship between load and temperature varies from region to region. There may be better variables to help with the load forecast than the peak and mean temperature.\\
- The inner relationship between energy use and different types of lead temperatures is investigated, significantly facilitating load forecast performance with the novel flexible temperature conditions designed in our paper.

\item Extreme weather events like wildfires raise the uncertainty and complexity of load patterns. Hence, the load behaviour analysis and forecast in regions encountering frequent extreme weather events are more worth studying than in the normal area \cite{AHMED2018911}.\\
- The wildfire seasons in 2015-2020 are chosen as the temporal scale of our study, filling the gap of load forecast under low-probability high-risk climate events.

\item While load forecasts at the DN level have been applied in some IEEE standardised networks, e.g., IEEE 33 bus model \cite{Morteza2021}, \cite{CHENG2021102856}, \cite{9909155}, the DN load forecast research based on real grids are relatively fewer conducted.\\
- Our paper developed the load forecast model based on authentic DN operations under extreme climates. Commonalities and differences in forecast performances in 8 real DNs are discussed and interpreted based on the regional multi-feature visualisation analysis.

\item Few studies have delved into the robustness of the load forecast model to extreme weather events. Research considering climate robustness is indispensable to propel progress in mitigating and adapting to Climate Change.\\
- Sensitivity and robustness analysis are conducted to prove the model reliability to data sets with large-scale climate variability.

\item Most existing research evaluated forecast performance solely based on technical metrics. However, the economic benefits calculated based on real network case studies can provide more intuitive and comprehensive guidance for system decision-makers and the authorities.\\
- A cost-benefit analysis is discussed to demonstrate the feasibility and sustainability of our model.
\end{itemize}

The rest of this paper is arranged as follows. Section 2 describes the multi-factor load forecast model's input matrix and main model structure. The forecast performance evaluation metrics, correlation analysis methods to form flexible temperature conditions, and techniques to visualise the data dispersion level are then discussed in detail. Section 3 introduces the background of our Case Study. Data sources and processing methods are explained for the original energy and climate input variables. Based on the forecast performance advancement, the generalised model is developed through four steps in Section 4, taking into consideration of input data structures, calendar labels and the flexible temperature condition assistance. In addition, robustness analysis and cost-benefit analysis are also discussed to prove model stability during extreme weather events and significant economic benefits. Section 5 summarises the key findings, advising  potential additional features that could strengthen the load forecast model in the future.

\section{Methodology} \label{method}
In this Section, the main steps to set up the multi-factor load forecast model are explained in detail. Then, the evaluation metrics are introduced to quantify the forecast performances. In addition, correlation analysis for energy demand and temperature are studied to explore how the climate factors affect the energy consumption behaviour so as to ascertain the optimal method that climate factors may be used to improve load forecast accuracy. Finally, methods to assess and visualise dispersion level of energy and climate data sets are discussed to help with the model generalisation step.

\subsection{Multi-factor deep learning forecast model}

\begin{figure}
	\centering
		\includegraphics[scale=.48]{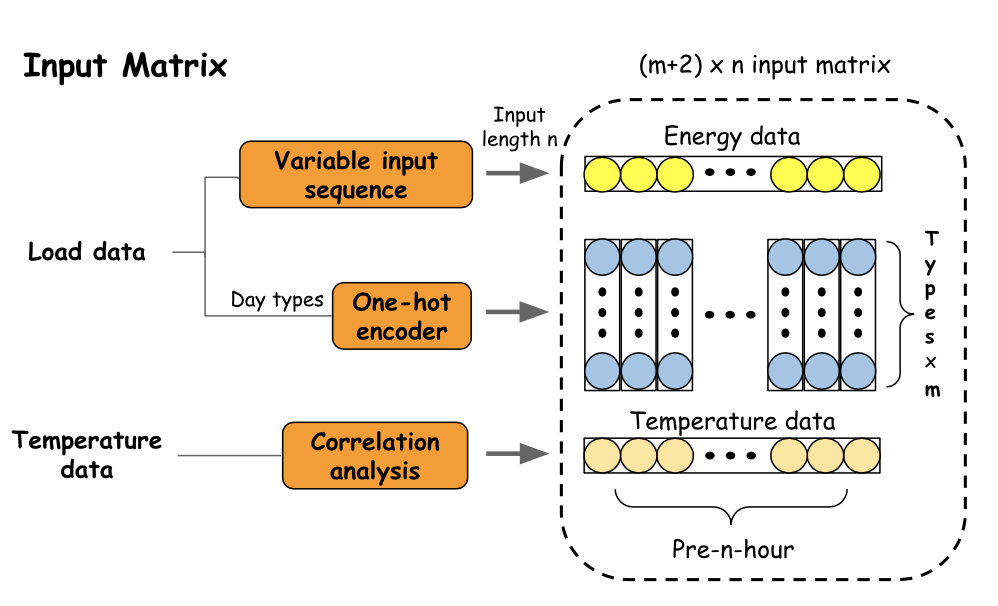}
	\caption{The structure of input matrix}
	\label{fig:Method_input}
\end{figure}

\begin{figure*}[htbp]
\begin{subfigure}{.45\textwidth}
\centering
\includegraphics[width=.96\columnwidth]{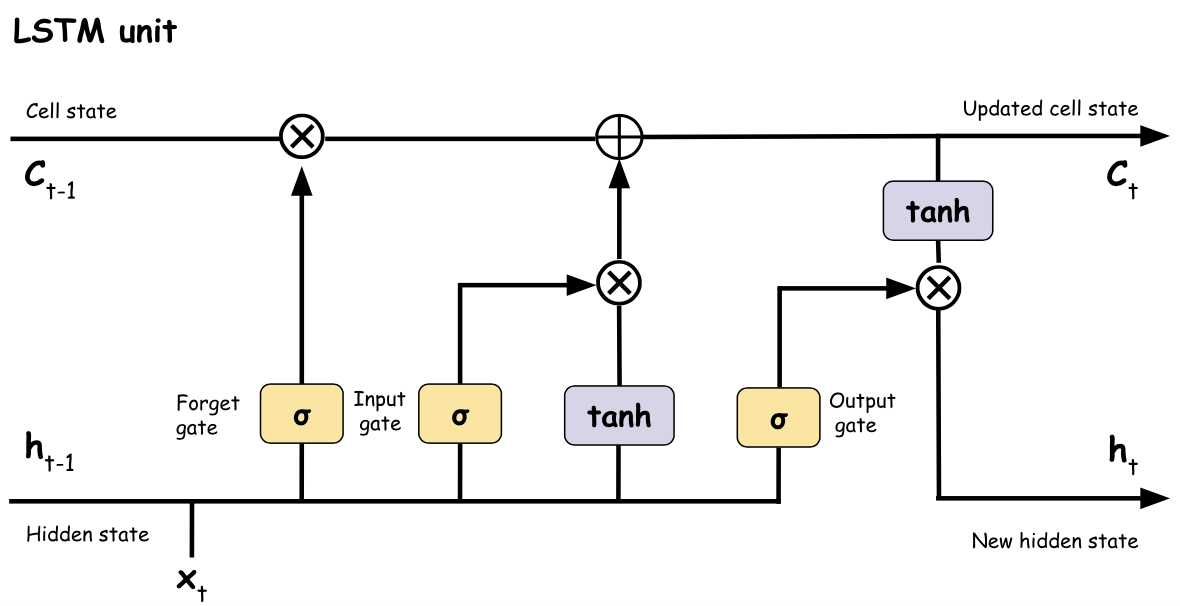}
\caption{LSTM cell architecture}
\label{fig:Method_LSTM}
\end{subfigure}
\hfill
\begin{subfigure}{.45\textwidth}
\centering
\includegraphics[width=0.98\columnwidth]{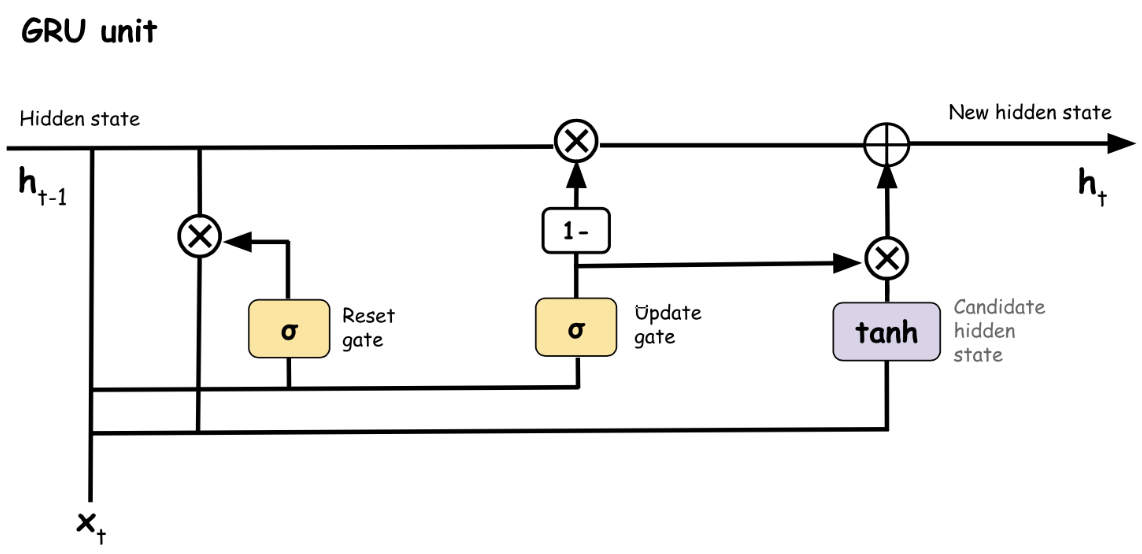}
\caption{GRU cell architecture}
\label{fig:Method_GRU}
\end{subfigure}
\caption{The structure of LSTM and GRU models}
\label{fig:Method_DL}
\end{figure*}

\begin{figure}
	\centering
		\includegraphics[scale=.48]{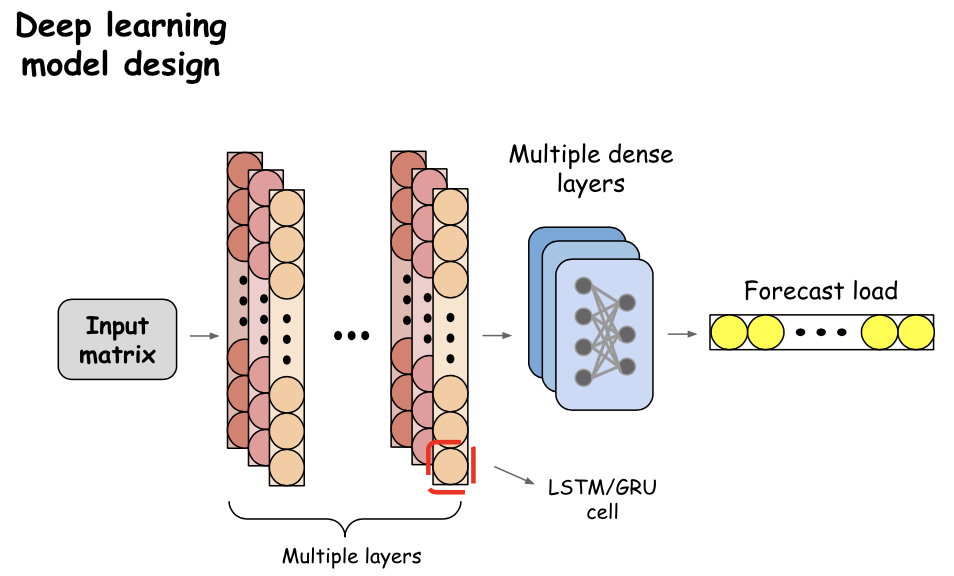}
	\caption{Multi-factor deep learning model design}
	\label{fig:Method_layer}
\end{figure}

In Figure~\ref{fig:Method_input}, the main steps to build the input matrix are explained. Load and temperature are the two main input sources. The load data set includes two features, one is the consumption data itself, and the other is the day information of each load data point. Different lengths of load input are tested in Step 1 to figure out the proper input length n with acceptable forecast accuracy and computational cost. Then, previous n load samples are set as the first row in the input matrix. The day information of load data points is classified into m types via a One-hot encoder to convert the categorical day type features to numeric variables \cite{Hancock2020}. The principle of One-hot encoding is as follows. A categorical variable \textbf{x} with n discrete values is expressed as $x_1, x_2,...,x_n$. Each category is transformed to a vector \textbf{v}, where every element of the vector is zero except for the $i^{th}$ which has the value of 1. For each load point, the day type vector is transposed to a column vector, and then day type information is arranged as a $m \times n$ matrix and placed in the middle of the input matrix. Temperature data sets are processed with correlation analysis to find the flexible temperature conditions at specific hours ahead that are strongly correlated with the load. Then, the proper conditional temperature data of each load point in pre-n-hours is placed in the last row. Thus, the input information is structured as a $(m+2) \times n$ matrix.\\
LSTM and GRU are the two deep learning models we have tested in our paper. LSTM and GRU are commonly used for forecasting with sequential input data. As load and climate data are time series, we consider LSTM and GRU to forecast load in our paper. As shown in Figure~\ref{fig:Method_LSTM}, each LSTM cell contains three main gates, the input gate, the forget gate and the output gate. New information is stored in the memory cell via the input gate. And then, part of the useful information is determined and remains in the current time step via the forget gate. Finally, the updated cell state and new hidden state information are output to generate the result and get ready for the next memory cell operation. $\sigma$ (sigmoid) and $tanh$ represent for activation functions to activate neurons accordingly and ensure the neural network can learn the non-linear behaviour and features \cite{Versloot2020}. Functional equations and weight matrix can be found in \cite{Eskandari2020} to further understand the LSTM cell architecture and theory.\\
As displayed in Figure~\ref{fig:Method_GRU}, GRU is a variant of LSTM with only two gates - the reset gate and the update gate. In 2014, GRU was proposed by \cite{cho-etal-2014-learning}. The cell and hidden states are combined as one state $h_{t}$. The reset gate determines how much past information should be forgotten in the current time step. The update gate generates the current hidden state $h_{t}$, using the previous hidden state $h_{t-1}$ and the candidate hidden state. Detailed equations and weight matrix formation can be studied in \cite{Eskandari2020} to further understand the GRU cell memory principle.\\
While LSTM and GRU both belong to RNN, they have advantages over each other in certain cases. GRU was invented later than LSTM, and the simplified gate structure uses less memory and is faster than LSTM. Also, it is proven that GRU structure is more immune to gradient exploding and vanishing issues \cite{César2021}. But for data sets with long and high-complexity sequences, LSTM has shown a better forecast performance \cite{Cahuantzi2021}. Thus, both LSTM and GRU structures are tested and compared in our paper.\\
The main structure of the LSTM and GRU deep learning models are displayed in Figure~\ref{fig:Method_layer}. First, the input matrix is put into the multiple LSTM/GRU layers, where the layer and neuron numbers can be set differently in various conditions. In general, the deterministic number of layers and nodes cannot be analytically calculated. There is a trade-off between forecast accuracy and computational efficiency: each data set has its optimal architecture. The appropriate perceptron structure can be explored by referring to existing deep learning models and adjusting with experimentation. Referring to the system in \cite{Rizk2021} and \cite{YANG2020117}, three LSTM/GRU layers with either 32 or 64 neurons are used to predict load with good accuracy and acceptable computational complexity in our paper. Then, multiple dense layers are connected to further process the forecast data. Dense layers are full connection layers with adjustable neurons. In our model, two dense layers are connected, with 16 and 1 neurons, respectively. Finally, the forecast load is output as a sequence. The forecast performance is evaluated on the testing data set.

\subsection{Forecast evaluation metrics}
In our paper, Mean Squared Error (MSE) is used to calculate the loss at each epoch to adjust the deep learning model training direction \cite{ZHANG20232705}. Also, MSE is used as one metric to evaluate the forecast performance in the \nameref{Results and Discussion}.
\begin{equation}
    MSE = \frac{1}{n}\ \sum_{i=1}^{n} (y-\hat{y})^{2}
\end{equation}
where: y is the actual value, $\hat{y}$ is the predicted value, and n is the number of data points.\\
Mean Absolute Percentage Error (MAPE) is another commonly used metric to assess forecast accuracy \cite{RUBASINGHE2023}. While MSE is a scale-based metric, the unit of MAPE is per cent, making it comparable among data sets with different units and scales.
\begin{equation}
    MAPE = \frac{1}{n}\ \sum_{i=1}^{n} |\frac{(y-\hat{y})}{y}|
\end{equation}
where: y is the actual value, $\hat{y}$ is the predicted value, and n is the number of data points.\\ 
MSE and MAPE help calculate the average error for the testing data set over the whole period. To assess the forecast performance diurnally, the average daily error plots with the temporal resolution of 30 minutes are discussed in the \nameref{Results and Discussion}. The daily error variance is computed to evaluate the forecast performance fluctuation within a day. The error variance $\sigma^{2}$ is calculated in Formula: 
\begin{equation}
    \sigma^{2} = \frac{\Sigma (X-\mu)^{2}}{N}\
\end{equation}
where: $\sigma^{2}$ is the error variance, X is the average half-hourly error, $\mu$ is the mean of the average half-hourly error over a day, and N is the total sampling points for a day. The stability of forecast accuracy within a day is worth studying to figure out the weak point of the forecast. The daily error variance is calculated to measure the model stability: the smaller the variance, the better robustness of the forecast model \cite{ZHENG2017}. And the total forecast model performance can be improved by assessing the hard-to-predict period in a day.

\subsection{Correlation analysis and data dispersion visualisation}

Correlation analysis is conducted to find the optimal temperature relationship that can improve the load forecast accuracy. For each DN, the Pearson Correlation Coefficient of the half-hourly load and the half-hourly instantaneous, maximum and average temperatures for lead time from 0 to 24 hours ahead are assessed to find the most strongly-correlated conditions.
\begin{equation}
    \rho_{X,Y} = \frac{cov{X,Y}}{\rho_{x}\rho_{y}}\
\end{equation}
where $\rho_{X,Y}$ is the Pearson Correlation Coefficient of X and Y, $cov$ stands for Covariance, $\rho_{x}$ and $\rho_{y}$ are the standard deviation of X and Y \cite{Saccenti2020}. In addition, the $1^{st}$ order $r^{2}$ value and the $2^{nd}$ order $r^{2}$ value are computed as other methods to study the relationship between load and leading temperatures.\\
To assess the dispersion level of temperature and load data sets at different DNs, each data set's Coefficient of Quartile Variation (CQV) and Standard Deviation (SD) are calculated and displayed in the data set visualisation plots in the \nameref{Results and Discussion}. As the CQV is a unitless metric, it is selected to measure the spread of data sets at different scales \cite{Yosboonruang2021}.
\begin{equation}
    CQV = \frac{Q3-Q1}{Q3+Q1}
\end{equation}
where $Q_{i}$ represents the $i^{th}$ quartile, and $Q_{3}$ - $Q_{1}$ is the interquartile range. In the data dispersion level visualisation plots, the $5^{th}$, $25^{th}$, $50^{th}$, $75^{th}$, and $95^{th}$ percentile lines are plotted to assist in analysing the daily data distribution.

\section{Case study}
This section describes the main steps of designing a generalised multi-factor deep learning load forecast model. Critical steps are explained in the context of the State of Victoria case study. Relevant data sources and processing methods for the load and temperature data used in our paper are presented in this section.

\subsection{Overview of the case study}
In Step 1, the input sequence length is varied to find a proper input data structure with acceptable low error and computational burden. Then, the performances of LSTM-based and GRU-based forecast models are compared, and the better one is selected as the main DL structure in the following steps. In Step 2, the effect of applying different calendar classification methods to input data is studied to improve the load forecast accuracy. In Step 3, correlation analysis is conducted to learn the relationship between temperature and load and to find the strongly-correlated temperature conditions. Some selected leading temperature conditions with high correlation coefficients are input into our model to improve the forecast performance as a novel method using flexible temperature conditions. In Step 4, we further apply the model to the other 7 DNs in Victoria, Australia, to study the regional difference and generalise our model.\\
The temperature tends to correlate stronger with load, playing a more essential role in load forecast under extreme weather conditions. Thus, regions with frequent extreme climate seasons are preferred to be the research region. Victoria, Australia, is selected as the research region for our paper due to its annual extreme wildfire season and data availability. In Step 3, one of the Victoria DNs, Horsham, is chosen to analyse the potential temperature function in improving the DN load forecast. As Horsham is located in the mid-latitude of Victoria State and has good data availability, we start with this location before generalising to other DNs in Victoria.

\subsection{Energy and climate data set acquisition and processing}

POWERCOR is one of the five main energy distributors in Victoria, Australia, covering western and central Victoria. As the mean summer wildfire risk (Forest Fire Danger Index) is generally higher in the western than in the eastern, POWERCOR coverage is selected as the research area in our paper \cite{BureauofMeteorology}. The energy consumption data set is downloaded through the Powercor Zone Substation Reports \cite{POWERCOR}. The load at different substations is recorded every 30 minutes in the POWERCOR database.\\
Temperature data is obtained from the Copernicus Climate data store. The 2m temperature netCDF data set is accessible from \cite{Copernicus}. The original temperature data has an hourly temporal resolution and 0.25\degree×0.25\degree spatial resolution. As the temporal resolution of load data is half-hourly, the temperature data set is temporally interpolated to half-hourly using Climate Data Operator (CDO) \cite{CDO} to keep it consistent with the load data set. In order to position the DN more precisely, the temperature data set is spatially interpolated into 0.01\degree×0.01\degree resolution, again using CDO. The temperature data of each DN is then extracted according to the city's geographical location at the spacial accuracy of 0.01\degree×0.01\degree. The consistent temporal resolution of input variables and a relatively high spatial resolution to locate the DN together contribute to a more accurate and smoother forecast process.

\section{Results and discussion} \label{Results and Discussion}
In this Section, we discuss the forecast performance enhancement at each step, determining the best design to develop the final generalised model. In Step 1, the effect of using different input data lengths is studied, and the forecast performances based on LSTM and GRU structures are compared for our case study. Various classifications of calendar labels are applied to the forecast model input to improve the sequential forecast accuracy in Step 2. And then, the relationship between temperature and load behaviour is discussed to explore flexible temperature conditions that may enhance the load forecast performance in Step 3. The operating cases with selected temperature conditions are tested over five wildfire seasons. The temperature condition that best contributes to the load forecast is determined based on three primary performance metrics. In Step 4, our forecast model is applied to 8 DNs in Victoria, forming a generalised and standardised forecast model.

\subsection{Step 1: Determining the optimal data input structure}
\begin{figure}
	\centering
		\includegraphics[width=.9\linewidth]{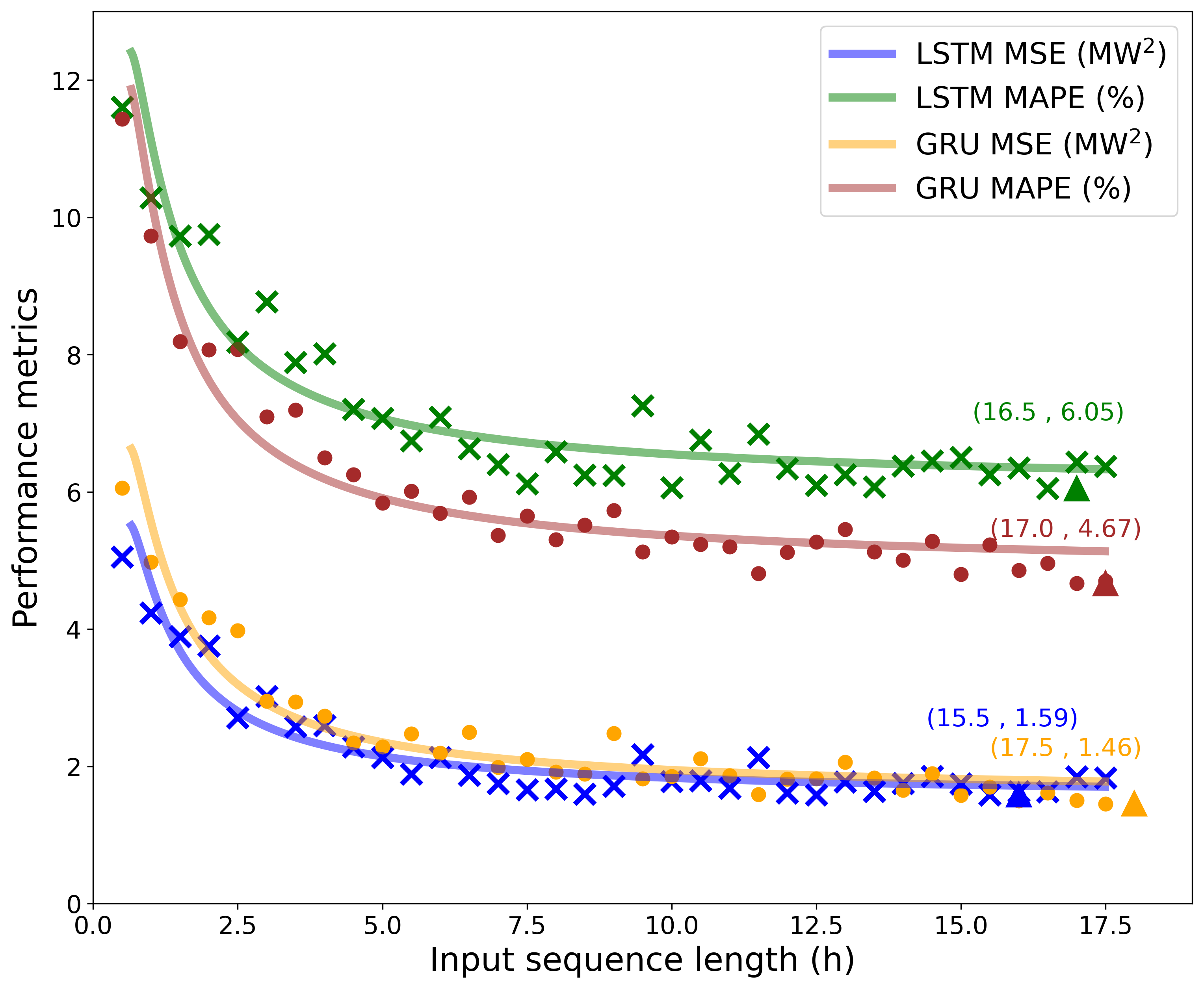} 
	\caption{Step 1 Horsham load forecast performance comparison during the wildfire seasons in 2015-2020 - different input sequence lengths and DL structures}
	\label{fig:Step_1}
\end{figure}

In the first step, the sequential energy consumption data set is input into LSTM and GRU models separately. The input data length is varied to observe the forecast error change, thus identifying the optimal data input length. In our paper, the input data length is increased by 30 minutes at every step since the temporal resolution of the Australian energy consumption data set is 30 minutes. As shown in Figure~\ref{fig:Step_1}, lines with crosses and dots represent the performances of LSTM-based and GRU-based models, respectively. And the points with the lowest error are marked with a triangle sign. MSE and MAPE decrease as the input length increases and reach a stable level when the input is around 16 hours for both LSTM and GRU cases.\\
It is demonstrated that the GRU-based model outperforms the LSTM-based model on MSE and MAPE by 8.18\% and 22.81\%, respectively. So, we primarily develop the model based on GRU structure and keep the data input length as pre-16-hour (32 sampling points) in the following steps. In later steps, the LSTM-based model is also tested as a comparison with results shown in the \nameref{Appen}.

\subsection{Step 2: Testing impact of calendar factors}
\begin{figure}
\centering
\begin{subfigure}[b]{1.0\linewidth}
\includegraphics[width=\linewidth]{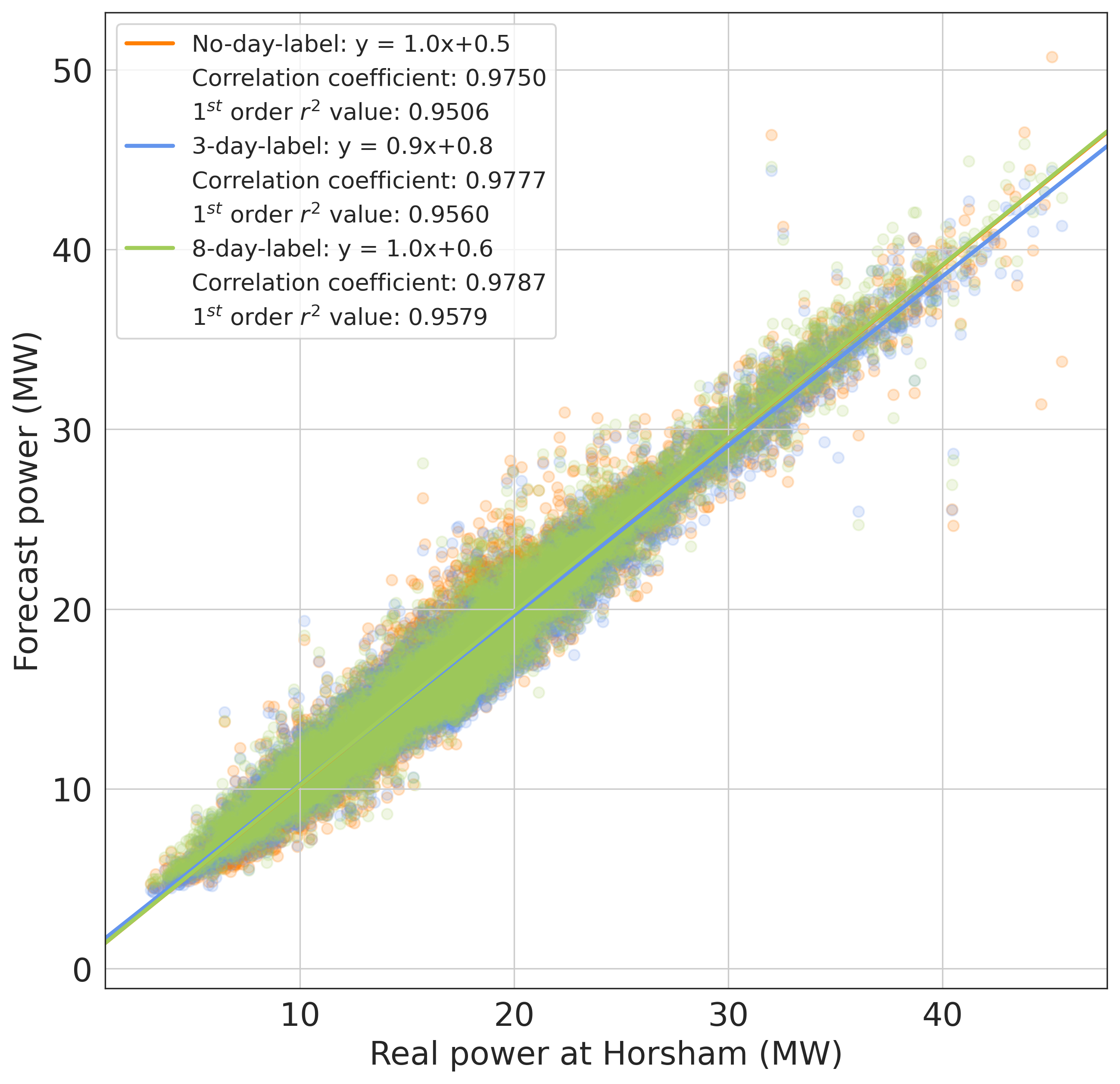}
\caption{Real power vs forecast power}
\label{fig:Step_2_1_a}
\end{subfigure}

\vspace{2ex}

\begin{subfigure}[b]{1.0\linewidth}
\includegraphics[width=\linewidth]{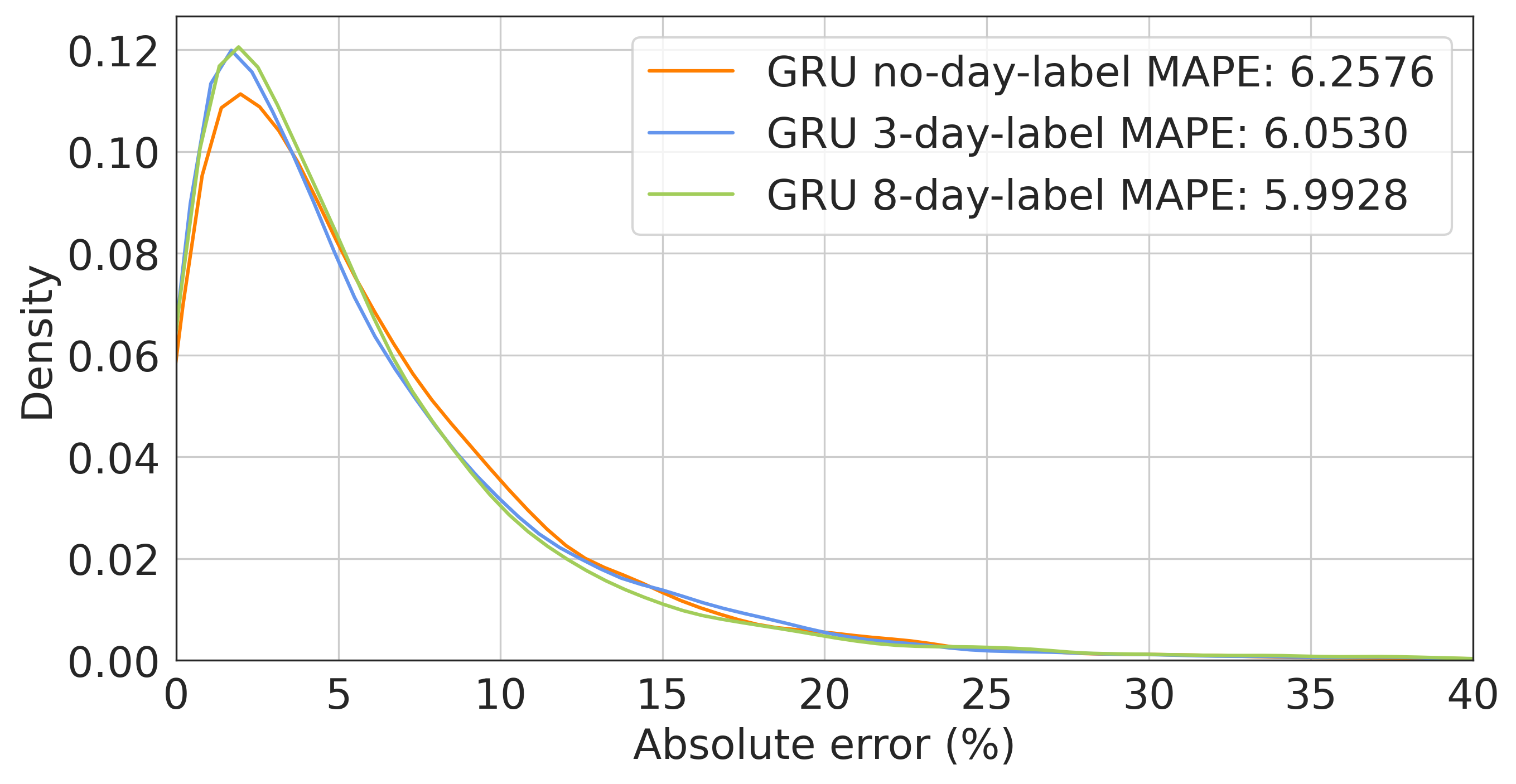}
\caption{Kernel Density Estimate - forecast error distribution}
\label{fig:Step_2_1_b}
\end{subfigure}

\vspace{2ex}

\begin{subfigure}[b]{1.0\linewidth}
\includegraphics[width=\linewidth]{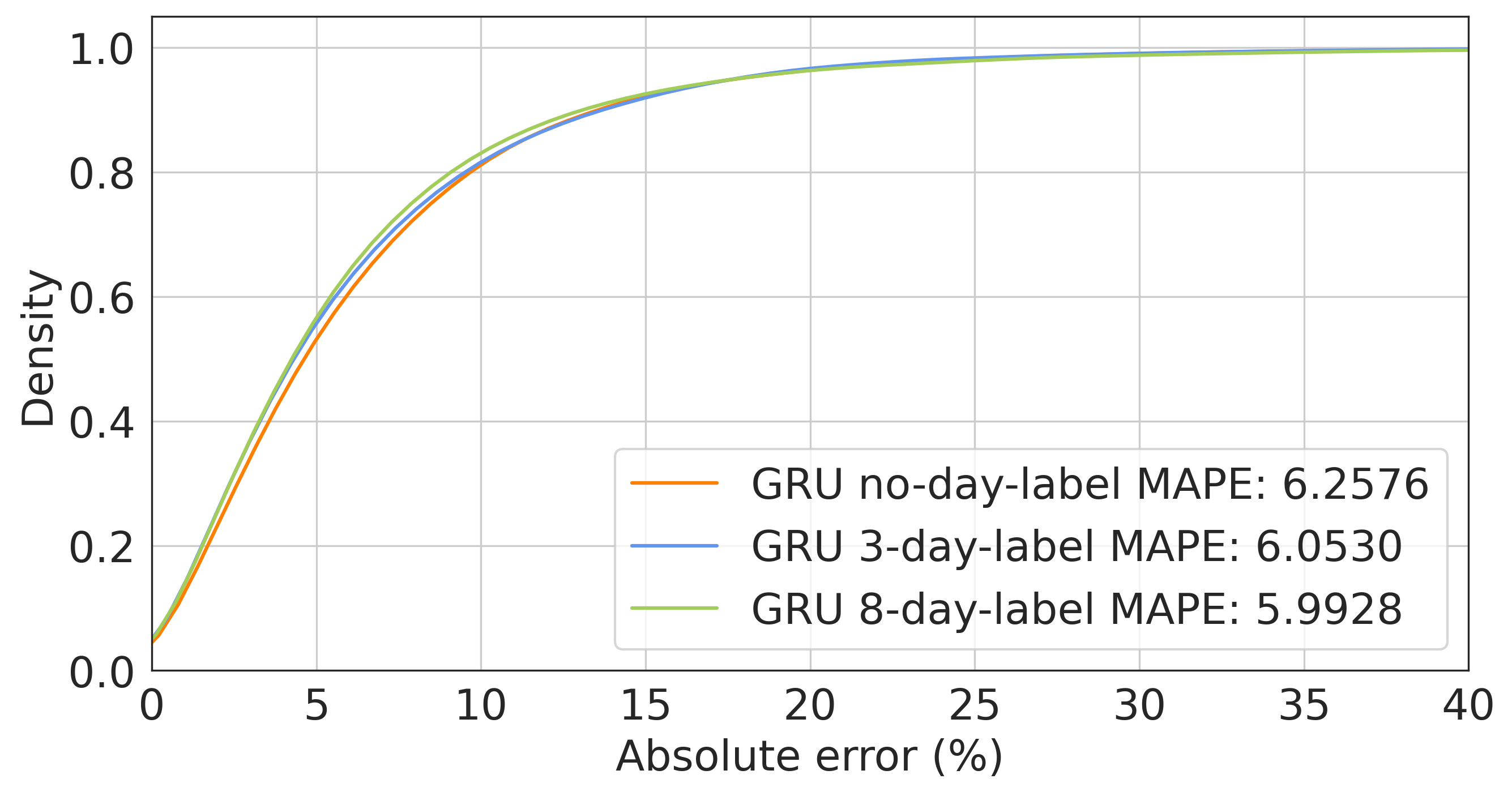}
\caption{Kernel Density Estimate Cumulative - forecast error distribution}
\label{fig:Step_2_1_c}
\end{subfigure}
\caption{Step 2 Horsham load forecast performance during the wildfire seasons in 2015-2020 - different calendar labels - GRU structure}
\label{fig:Step_2_1}
\end{figure}

\begin{figure}[hbt!]
\centering
\begin{subfigure}[b]{1.0\linewidth}
\includegraphics[width=\linewidth]{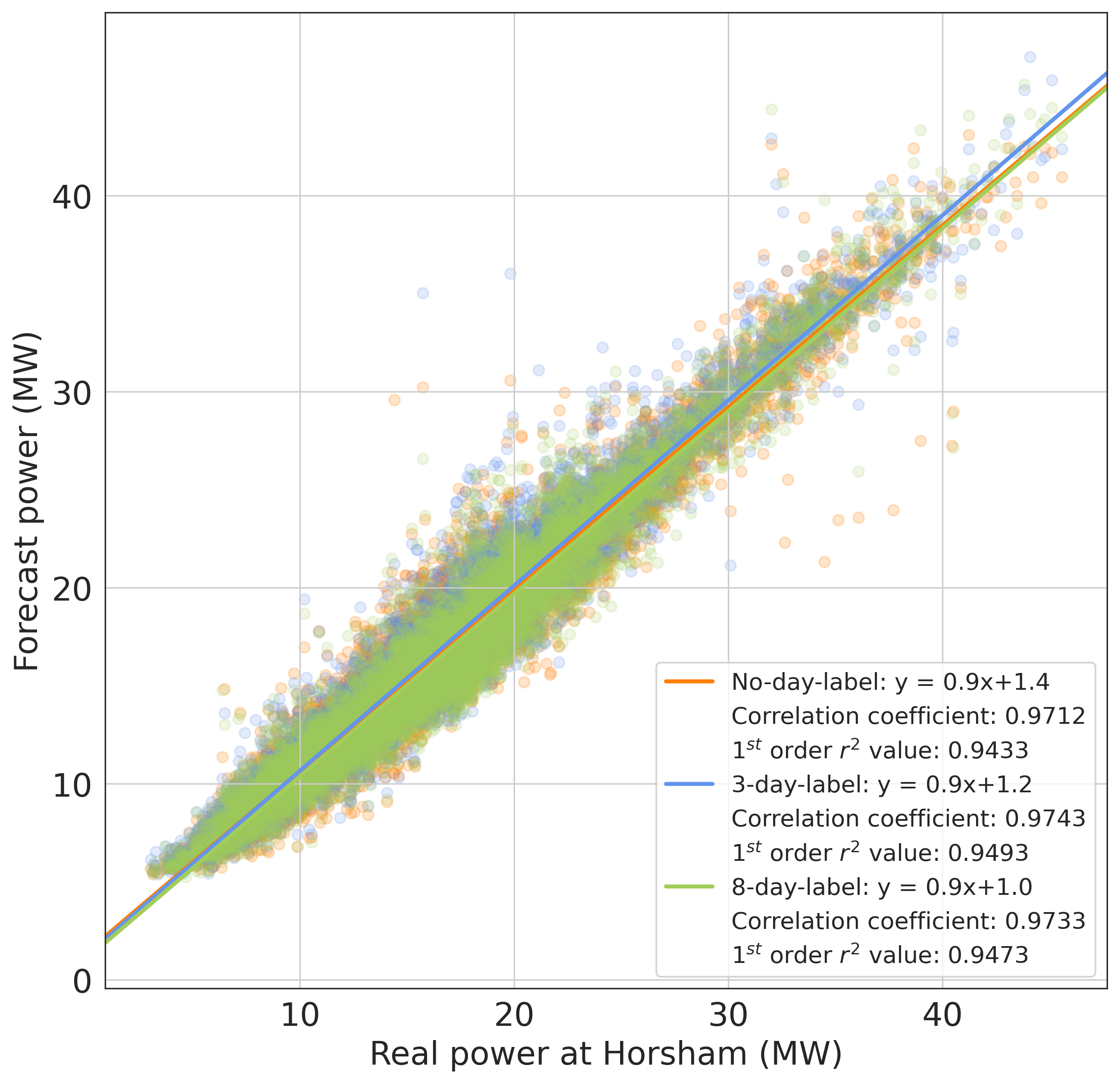}
\caption{Real power vs forecast power}
\label{fig:Appendix_1_a}
\end{subfigure}

\vspace{2ex}

\begin{subfigure}[b]{1.0\linewidth}
\includegraphics[width=\linewidth]{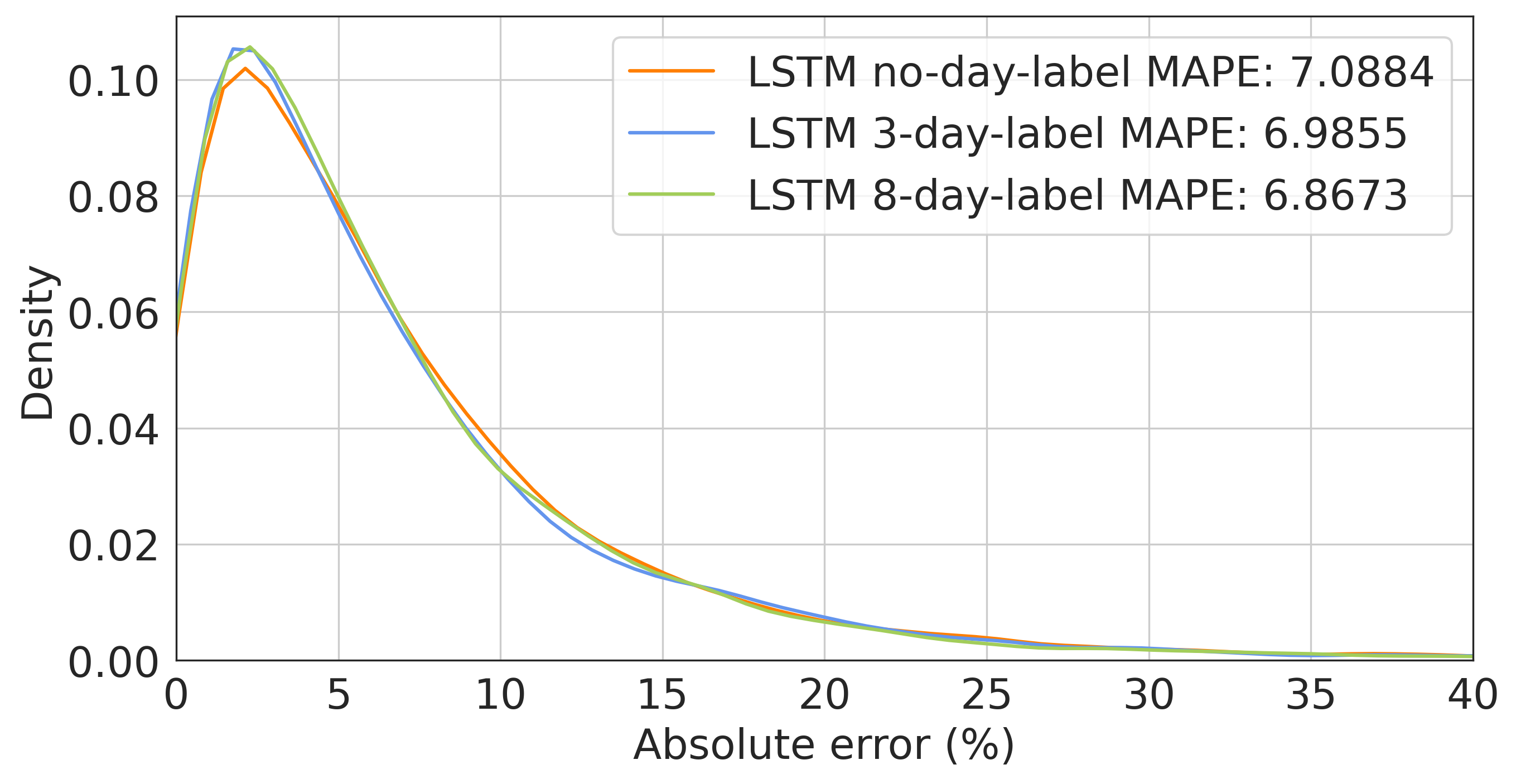}
\caption{Kernel Density Estimate - forecast error distribution}
\label{fig:Appendix_b}
\end{subfigure}
\vspace{2ex}

\begin{subfigure}[b]{1.0\linewidth}
\includegraphics[width=\linewidth]{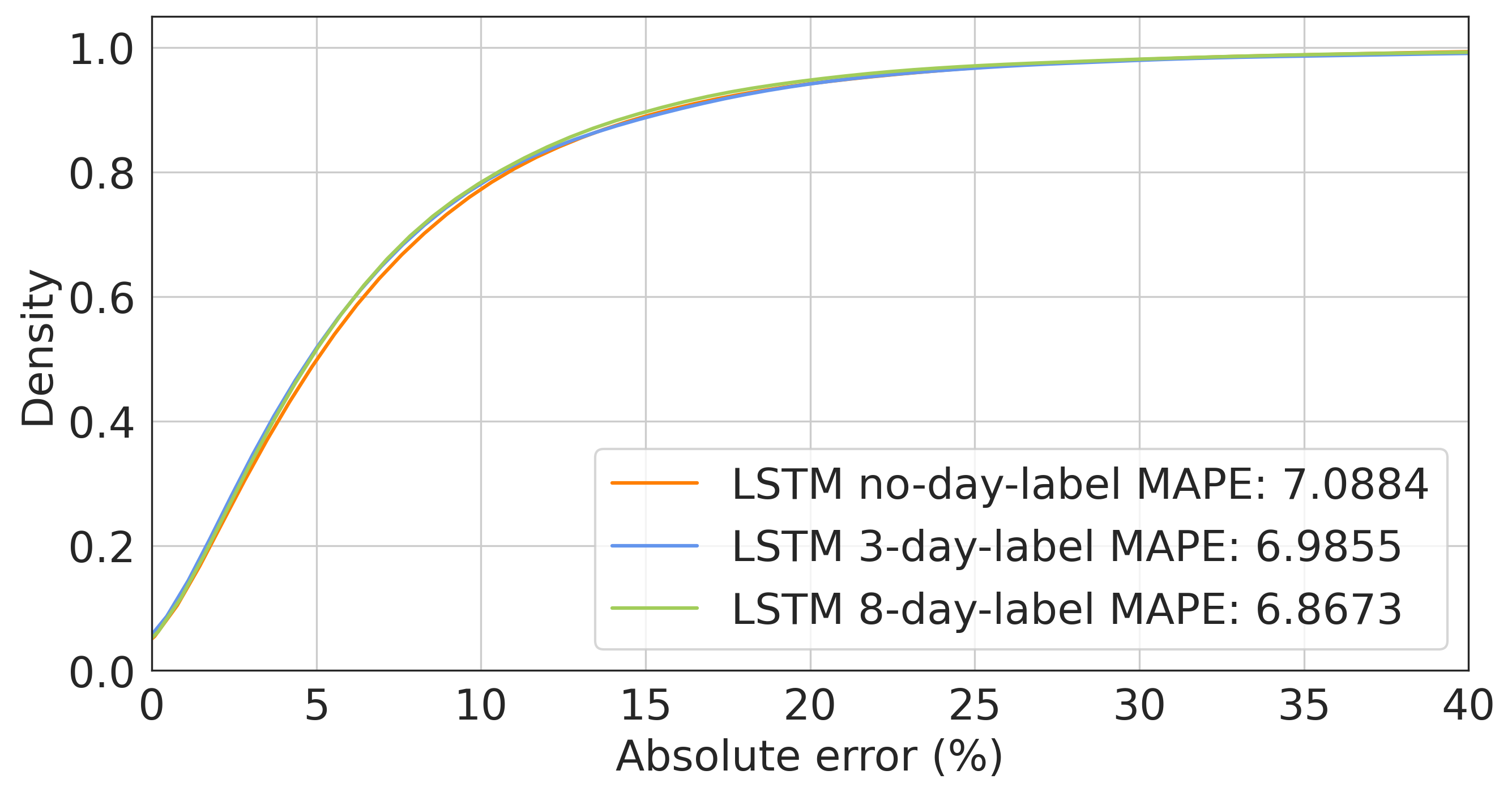}
\caption{Kernel Density Estimate Cumulative - forecast error distribution}
\label{fig:Appendix_1_c}
\end{subfigure}
\caption{Step 2 Horsham load forecast performance during the wildfire seasons in 2015-2020 - different calendar labels - LSTM structure}
\label{fig:Appendix_1}
\end{figure}

The second step studies the effect of considering calendar factors in our model input. In this step, three operating conditions are tested: no-day-type-label, three-day-type-label, and eight-day-type-label. For the three-day type label conditions, the day types are classified into weekday, weekend, and holiday. The eight-day-type-label conditions are classified into Monday to Sunday, plus a holiday label. As a comparison group, the no-day-type-label condition is tested with the original energy consumption data as the only input.\\
In Figure~\ref{fig:Step_2_1_a}, the real power versus three forecast powers using the GRU-based model during the testing period (the wildfire seasons in 2018-2020) are visualised and compared. The three forecast power are strongly correlated with the real power. Among all three cases, the two with calendar labels perform better than the one without day-type tags on the correlation coefficient, especially at energy consumption peaks. Quantitatively, the MAPE and the error Kernel Density Estimate (KDE) plots of the three forecast cases are displayed in Figure~\ref{fig:Step_2_1_b} and Figure~\ref{fig:Step_2_1_c}. The MAPE of the eight-day-type-label case is 4.23\% lower than not using the day-type-label and 1.01\% lower than the three-day-type-label case. In the KDE distribution and cumulative plots, it is demonstrated that the majority of error points for the eight-day-type-label case gather at a lower value than the other two cases.\\
Similar to the findings from Step 1, GRU still outperforms LSTM when using the same data input matrix: the eight-day-type-label MSE and MAPE of the GRU-based model are 20.4\% and 12.73\% lower than the LSTM model with the eight-day-type-label, respectively. Among various day-label use cases, the lowest forecast error of the LSTM-based model is also the one with the eight-day-type-label, which again verifies that the eight-day-type-label is more helpful than the three-day-type-label for our case study. Thus, we continue to use the pre-16-hour input \& eight-day-type-label GRU model as the main structure in the following steps. The correlation analysis and error distribution plots for the LSTM case are displayed in Figure~\ref{fig:Appendix_1} in the \nameref{Appen} as a reference.

\subsection{Step 3: Designing flexible temperature conditions to improve load forecast}
\begin{figure}[hbt!]
\centering
\begin{subfigure}[b]{1.0\linewidth}
\includegraphics[width=\linewidth]{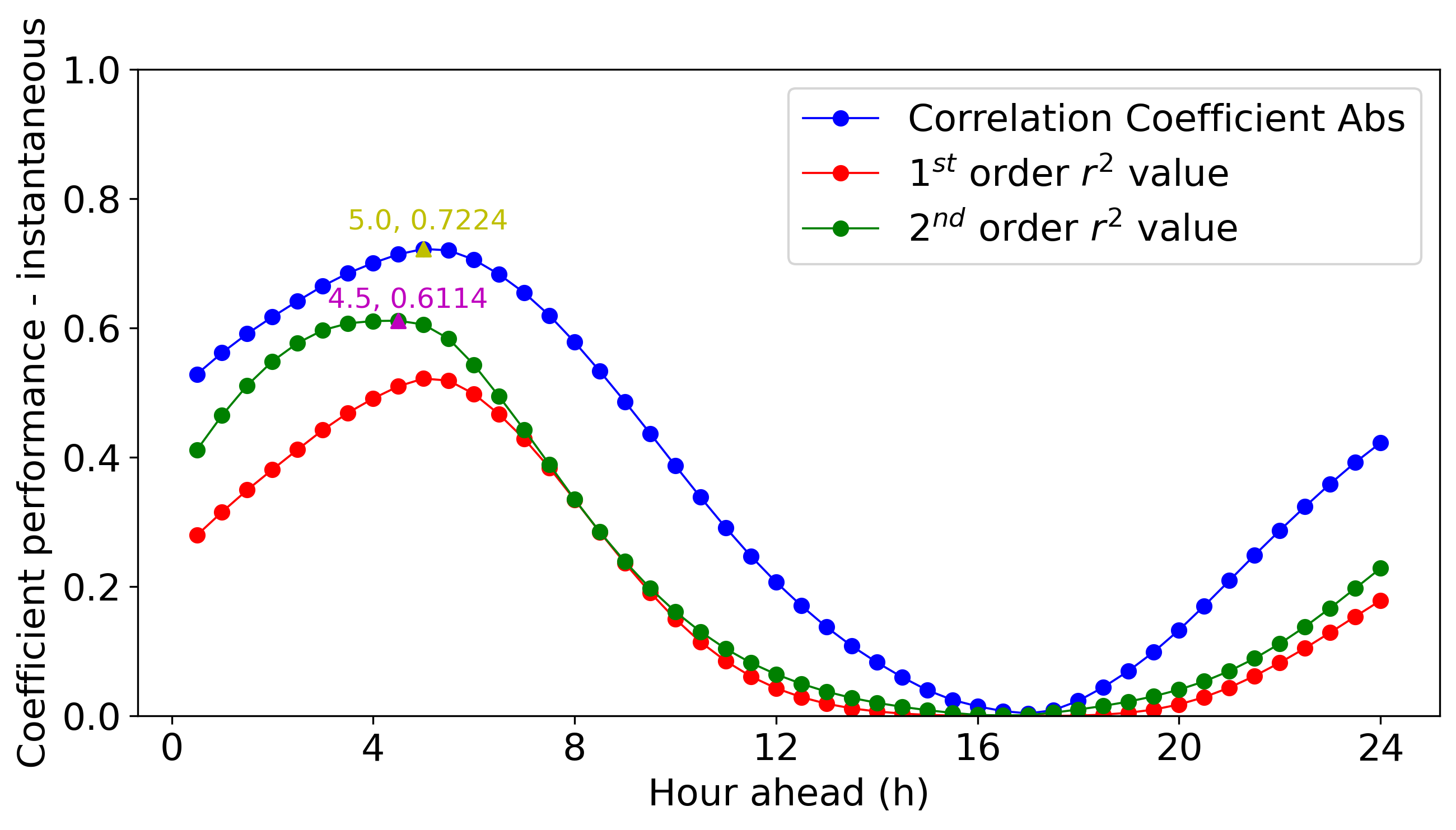}
\caption{Instantaneous temperature}
\label{fig:Step_3_1_a}
\end{subfigure}

\vspace{2ex}

\begin{subfigure}[b]{1.0\linewidth}
\includegraphics[width=\linewidth]{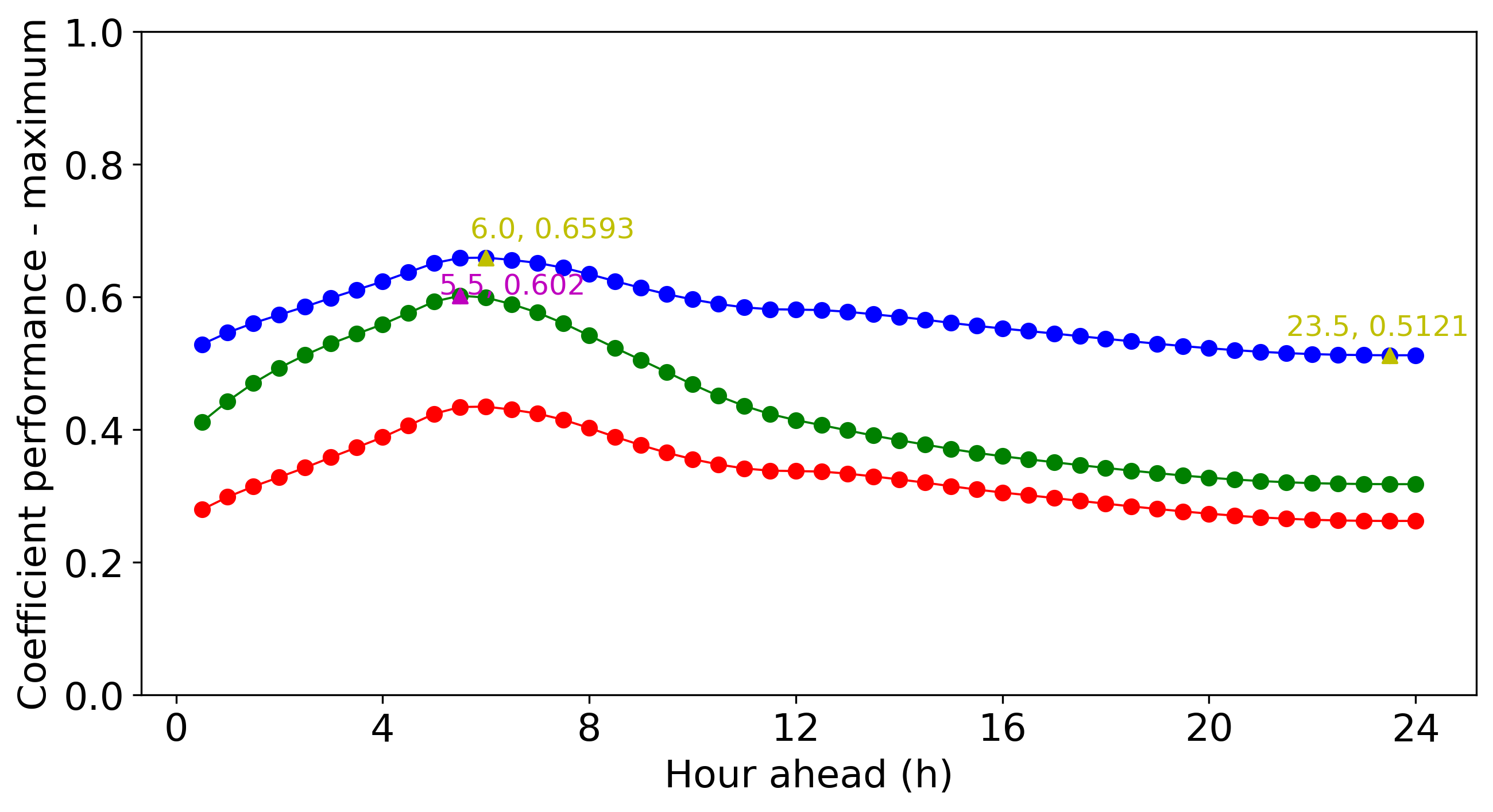}
\caption{Maximum temperature}
\label{fig:Step_3_1_b}
\end{subfigure}

\vspace{2ex}

\begin{subfigure}[b]{1.0\linewidth}
\includegraphics[width=\linewidth]{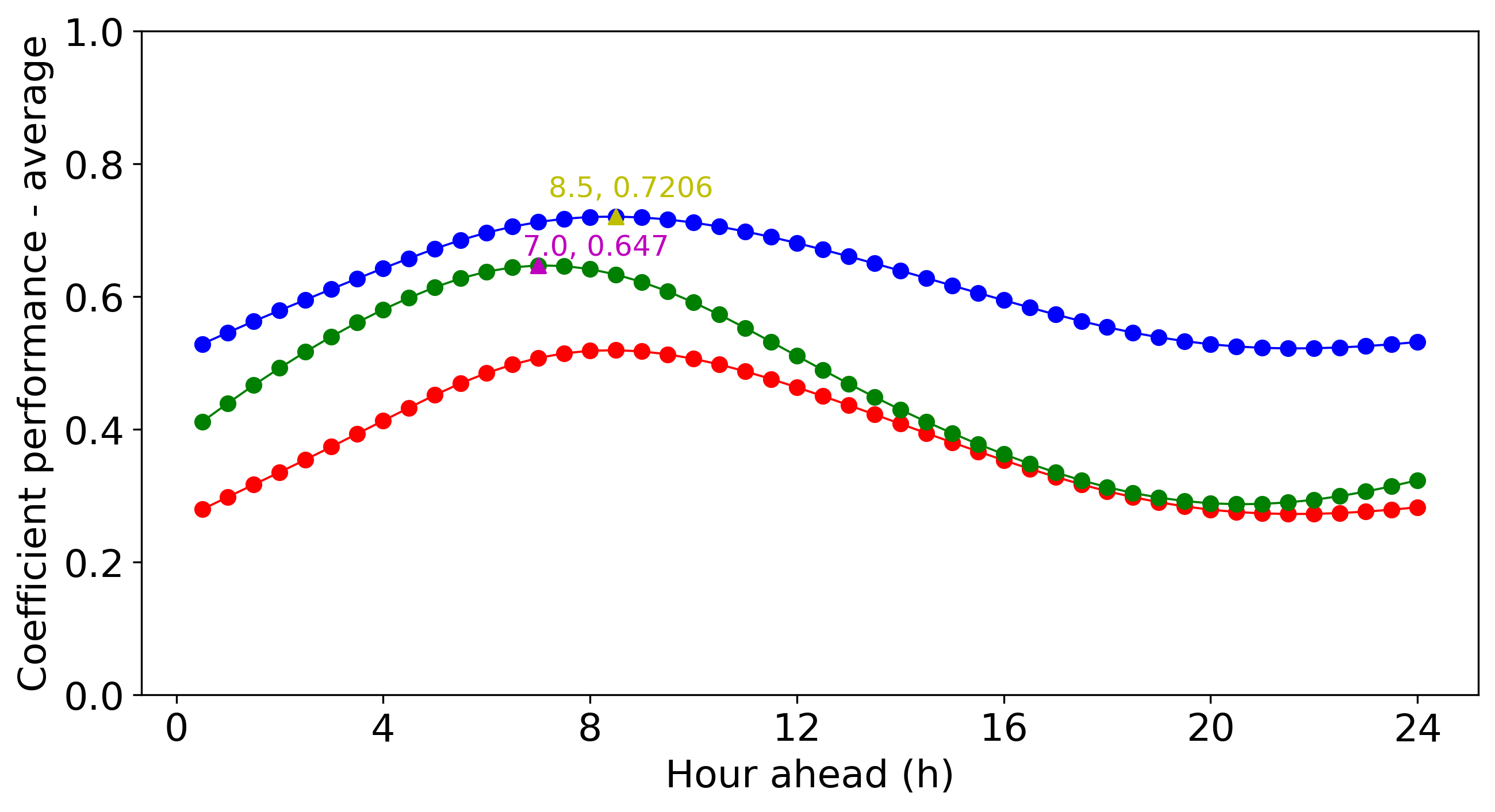}
\caption{Average temperature}
\label{fig:Step_3_1_c}
\end{subfigure}
\caption{Step 3 correlation analysis for load and various preceding temperatures during the wildfire seasons in 2015-2020 in Horsham, Victoria, Australia}
\label{fig:Step_3_1}
\end{figure}

In Step 3, the relationship between energy consumption and temperature is first studied. It is postulated that the energy consumption behaviour on the DN level is affected by the preceding instantaneous, average, and maximum temperatures at different timescales. We therefore consider the optimal temperature relationship in the deep learning model to forecast load with the help of flexible temperature conditions.\\
There is some existing research on energy forecast quality and its dependence on temperature. However, most existing papers select daily mean temperature or instantaneous temperature as the auxiliary input without giving a reason \cite{6140142}, \cite{Zeng2018}, \cite{MALDONADO2019105616}, \cite{Wang2021IEEE}. In our paper, we implement pre-n-hour instantaneous, maximum and average temperature correlation analyses with energy consumption, ranging from 0.5 to 24 hours ahead. Only the temperature-leading-load situation is assessed in our paper since the real-time temperature change can lead the energy consumption change. In contrast, temperature is a natural parameter that cannot be affected by energy consumption over a short scale. In Figure~\ref{fig:Step_3_1}, the preceding n-hour instantaneous, maximum and average temperatures are tested at increments of 0.5 hours, from 0 to 24-hour leading. The correlation coefficient, the $1^{st}$ order $r^{2}$ value and the $2^{nd}$ order $r^{2}$ value are plotted for load and temperature. As postulated, the load is strongly correlated with the temperature ahead: the highest correlation coefficient is greater than 0.7 for different lead times of instantaneous and average temperatures. The $2^{nd}$ order $r^{2}$ value curve demonstrates the same trend as the correlation coefficient and peaks at 0.64 under the pre-7-hour average temperature condition. The temperature conditions that lead to the strongest correlation and the best $2^{nd}$ order $r^{2}$ value for three temperature types are recorded in Table~\ref{tbl1}.

\begin{table*}[width=2.06\linewidth,cols=7]
\caption{The strongest correlation conditions for instantaneous, maximum, and average temperature in Horsham, Victoria, Australia (the wildfire seasons in 2015-2020)}\label{tbl1}
\begin{tabular*}{\tblwidth}{@{} LLLLLLL@{} }
\toprule
Conditions & \multicolumn{2}{c}{Instantaneous temp} & \multicolumn{2}{c}{Maximum temp} & \multicolumn{2}{c}{Average temp}\\
\midrule
Metrics & Hour & Value & Hour & Value & Hour & Value \\
Correlation coefficient & 5hr lead & 0.721 & 6hr lead & 0.6581 & 8.5hr lead & 0.7192 \\
$2^{nd}$ order $r^{2}$ value & 4.5hr lead & 0.6086 & 5.5hr lead & 0.5995 & 7hr lead & 0.6438 \\
\bottomrule
\end{tabular*}
\end{table*}

\begin{figure}[hbt!]
\centering
\begin{subfigure}[b]{1.0\linewidth}
\includegraphics[width=\linewidth]{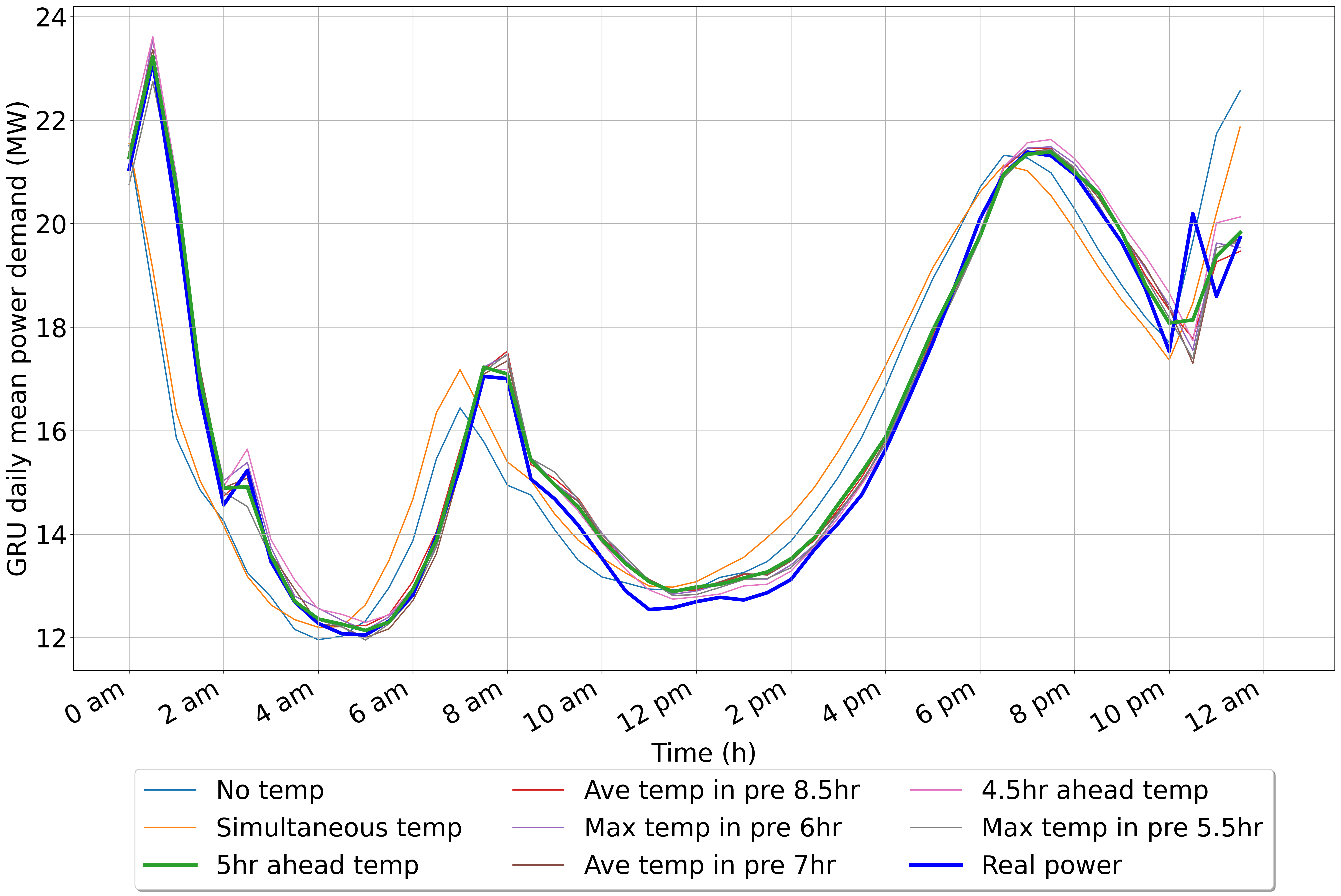}
\caption{Daily mean load \& forecast load}
\label{fig:Step_3_2_a}
\end{subfigure}

\vspace{2ex}

\begin{subfigure}[b]{1.0\linewidth}
\includegraphics[width=\linewidth]{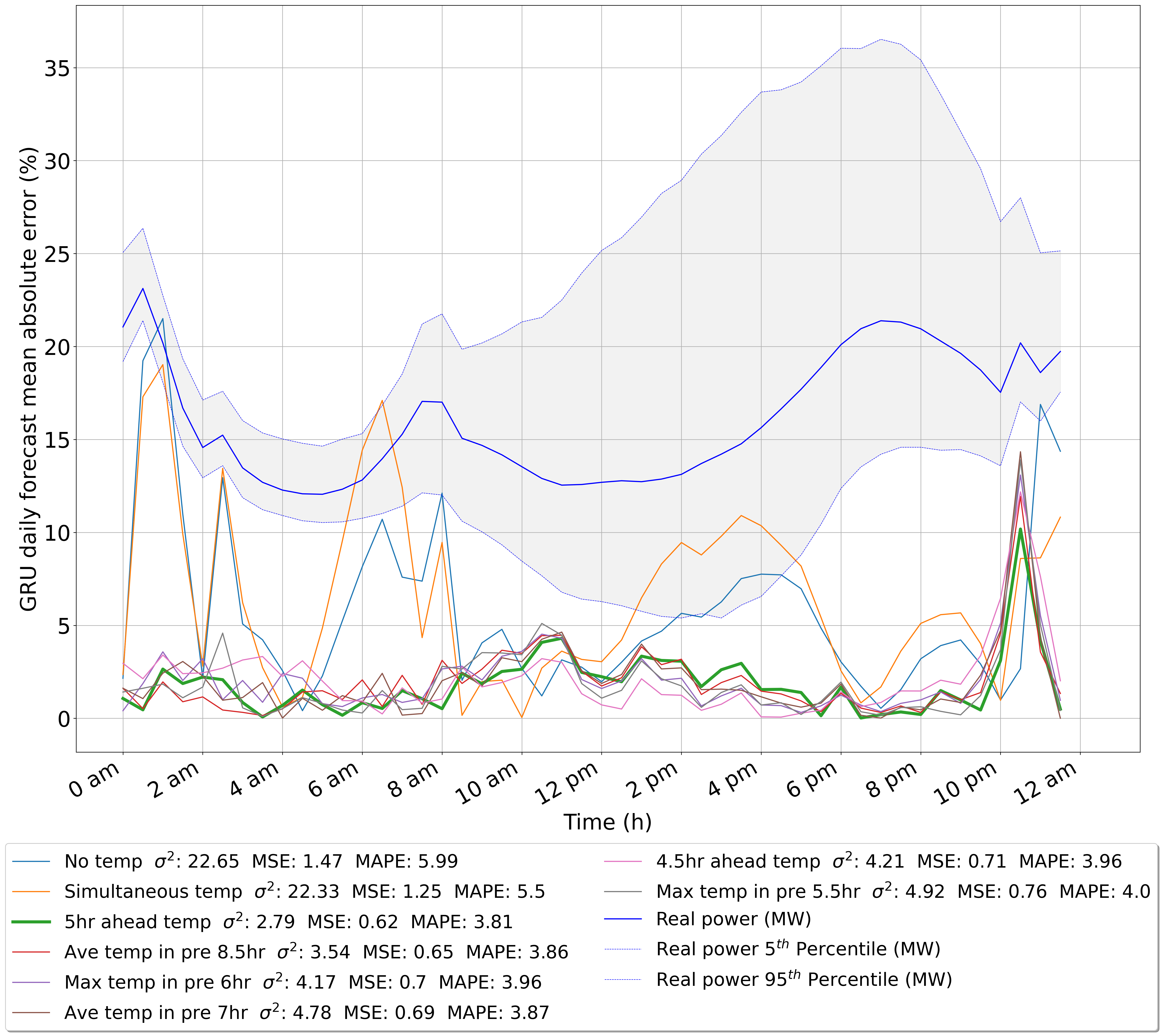}
\caption{Daily forecast mean errors with different temperature uses}
\label{fig:Step_3_2_b}
\end{subfigure}
\caption{Step 3 Horsham load forecast performance during the wildfire seasons in 2015-2020 - flexible temperature conditions impact study - GRU structure}
\label{fig:Step_3_2}
\end{figure}

\begin{figure}[hbt!]
\centering
\begin{subfigure}[b]{1.0\linewidth}
\includegraphics[width=\linewidth]{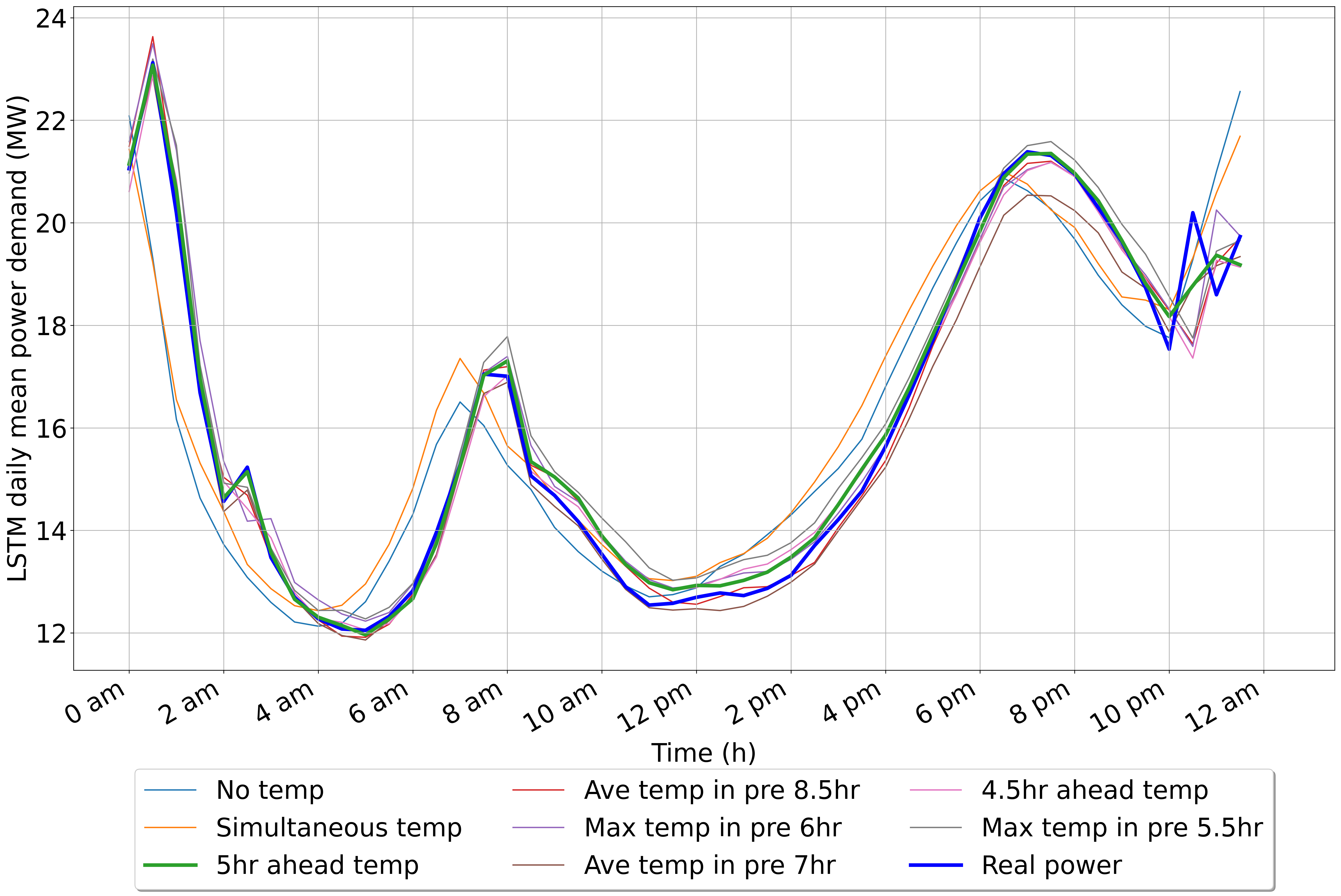}
\caption{Daily mean load \& forecast load}
\label{fig:Appendix_2_a}
\end{subfigure}

\vspace{2ex}

\begin{subfigure}[b]{1.0\linewidth}
\includegraphics[width=\linewidth]{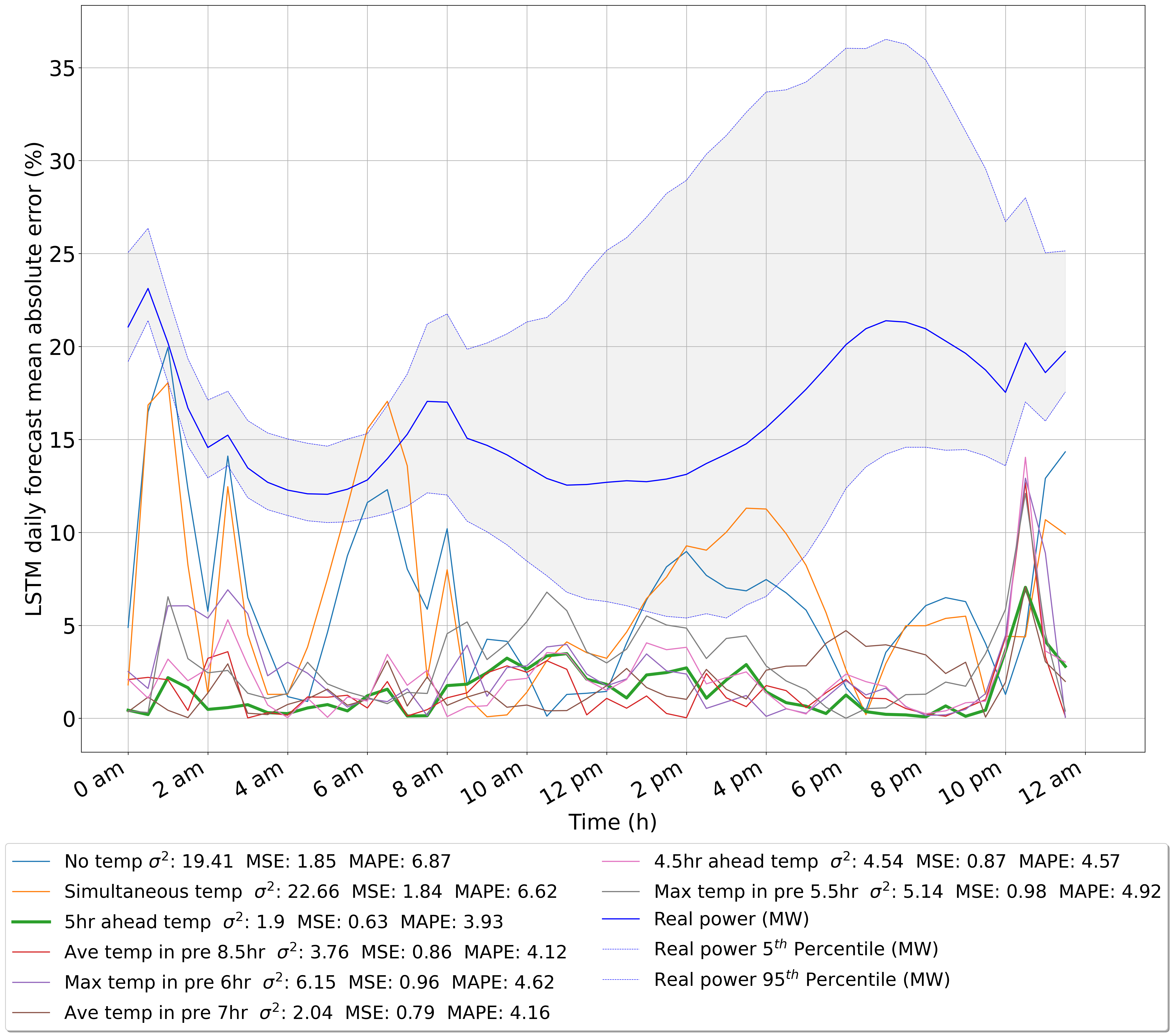}
\caption{Daily forecast mean errors with different temperature uses}
\label{Appendix_2_b}
\end{subfigure}
\caption{Step 3 Horsham load forecast performance during the wildfire seasons in 2015-2020 - flexible temperature conditions impact study - LSTM structure}
\label{fig:Appendix_2}
\end{figure}

In Step 3, the six flexible temperature conditions in Table~\ref{tbl1} are input into our model to improve load forecast accuracy. As comparison groups, the model with the eight-day-type-label but no temperature (from Step 2) and the model using simultaneous-load-temperature are also plotted to study the positive effect of adding leading temperature into our model. As shown in Figure~\ref{fig:Step_3_2_a}, eight forecast load curves describe a trend similar to the actual daily average load during the wildfire seasons in 2015-2020. The bold blue line represents the real daily average load. The 5-hour-ahead instantaneous temperature case is displayed with a bold green line since it performs the best among all eight forecast conditions. Noticeably, there is a horizontal deviation between the real load curve and the two forecast lines without using the assistant flexible temperature conditions from Table~\ref{tbl1} (the light blue and orange lines). After applying leading temperature conditions to the input matrix, the forecast lines pan right and move towards the real load.\\
In Figure~\ref{fig:Step_3_2_b}, the forecast performance is quantified to study the temperature condition that best improves the forecast accuracy. The blue line shows the real daily load, and the grey area surrounded by the blue dotted line represents the $5^{th}$ to $95^{th}$ percentile interval of the daily load in the wildfire seasons in 2015-2020. It is common that the daily average load has two peaks around 8 am and 7 pm, the so-called the "Duck Curve" \cite{HOU2019205}. The additional uptick in consumption noticeable between 10:30 pm and 1:30 am is caused by dedicated loads from off-peak electrical hot water systems \cite{SustainabilityVic2022}. The grey area is wider during the day and more densely distributed during the night, indicating the load dispersion changes across the day. In terms of the absolute daily error, the half-hourly error is mostly controlled within 5\% for the six temperature-leading conditions, while the two comparison cases without using leading temperature demonstrate errors of up to 20\%. The green line (5-hour-ahead instantaneous temperature) is marked bold to highlight the best performance condition in this step: the error variation is 2.79$MW^{2}$, MSE is 0.62$MW^{2}$, and MAPE is 3.81\%. Compared to the simultaneous-temperature-load and the no-temperature methods, the MAPE with the proposed best correlation coefficient temperature condition is decreased by 30.73\% and 36.39\%, respectively. We have also tested the forecast performance when including the eight conditions with the LSTM model. The performance result plots are demonstrated in Figure~\ref{fig:Appendix_2} in the \nameref{Appen}. The GRU based forecast model continues outperforming that using LSTM in MSE and MAPE by 1.59\% and 3.05\% in the best-correlated temperature condition case in Step 3.

In this step, we have identified the optimal way to utilise, as additional inputs, the temperature conditions to provide increased accuracy of the energy consumption forecast. The pre-n-hour correlation analysis for instantaneous, maximum and average temperature conditions is first implemented, showing the strong correlation between leading temperature and DN load behaviour. The best correlation coefficient conditions and the best $2^{nd}$ order $r^{2}$ value conditions are defined as additional inputs to the model. The best correlation coefficient GRU-based case stands out among all cases. Therefore, we continue to apply the best correlation coefficient temperature condition with the GRU structure in the next model generalisation step.

\subsection{Step 4: Model generalisation and standardisation}
\begin{figure}
	\centering
		\includegraphics[width=1.0\linewidth]{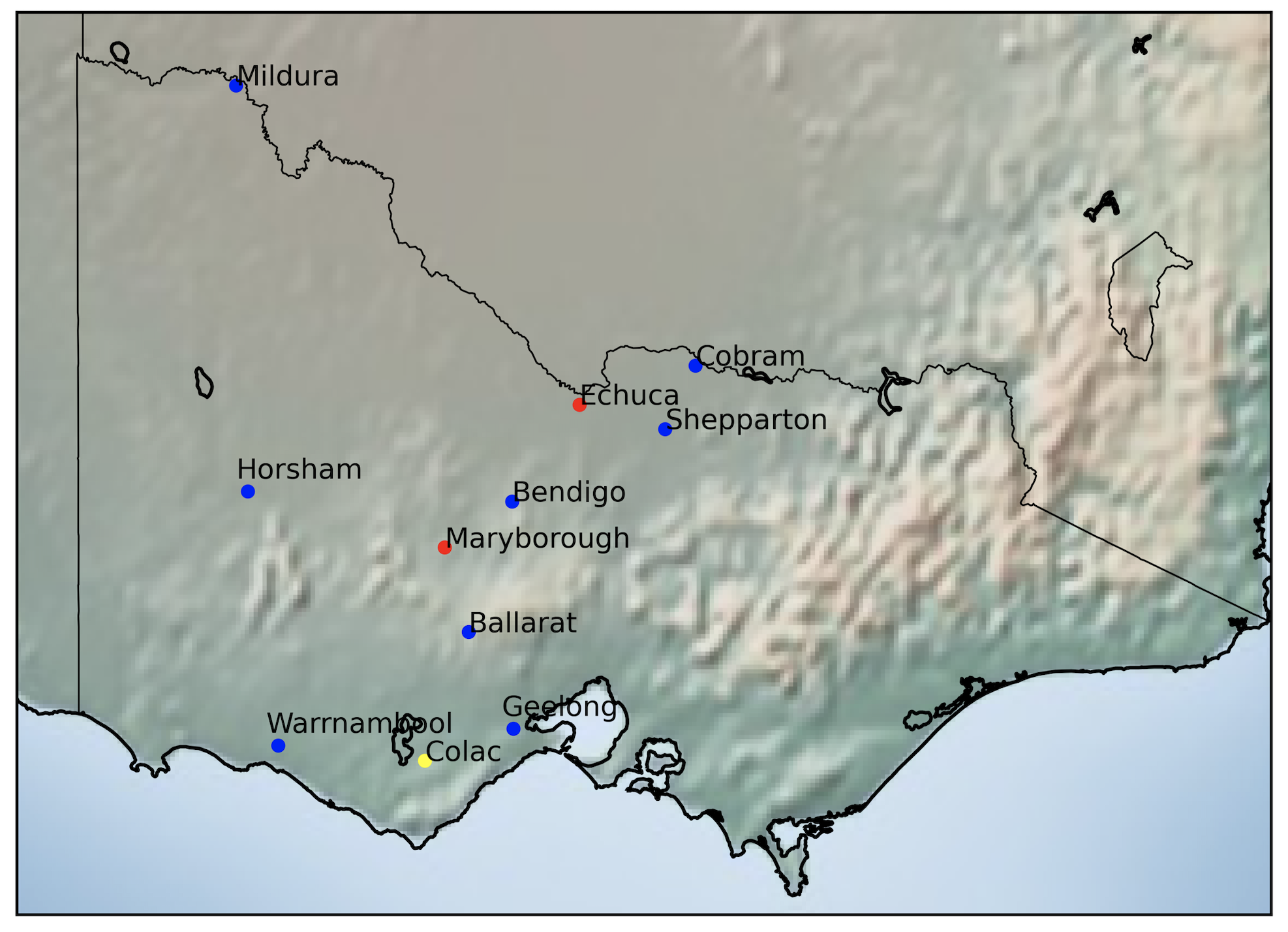} 
	\caption{Map of 11 distribution networks in Victoria, Australia (blue: selected, red: poor data quality, yellow: insensitive to temperature)}
	\label{fig:Step_4_1}
\end{figure}

In this step, our model is applied to other DNs in Victoria, Australia, to study possible regional differences and further improve the model. As shown in Figure~\ref{fig:Step_4_1}, there are eleven zone substations in the POWERCOR-owned area in Victoria. Echuca and Maryborough are marked with red dots due to poor data quality. Colac is represented with a yellow dot indicating the load is insensitive to temperature change. As our paper aims to discuss the advantage of using temperature in load forecasting, we do not look at temperature-insensitive locations such as Colac. The other eight DN are marked with blue dots, and these eight regional loads are tested in this step.

\begin{figure}[hbt!]
\centering
\begin{subfigure}[b]{1.0\linewidth}
\includegraphics[width=\linewidth]{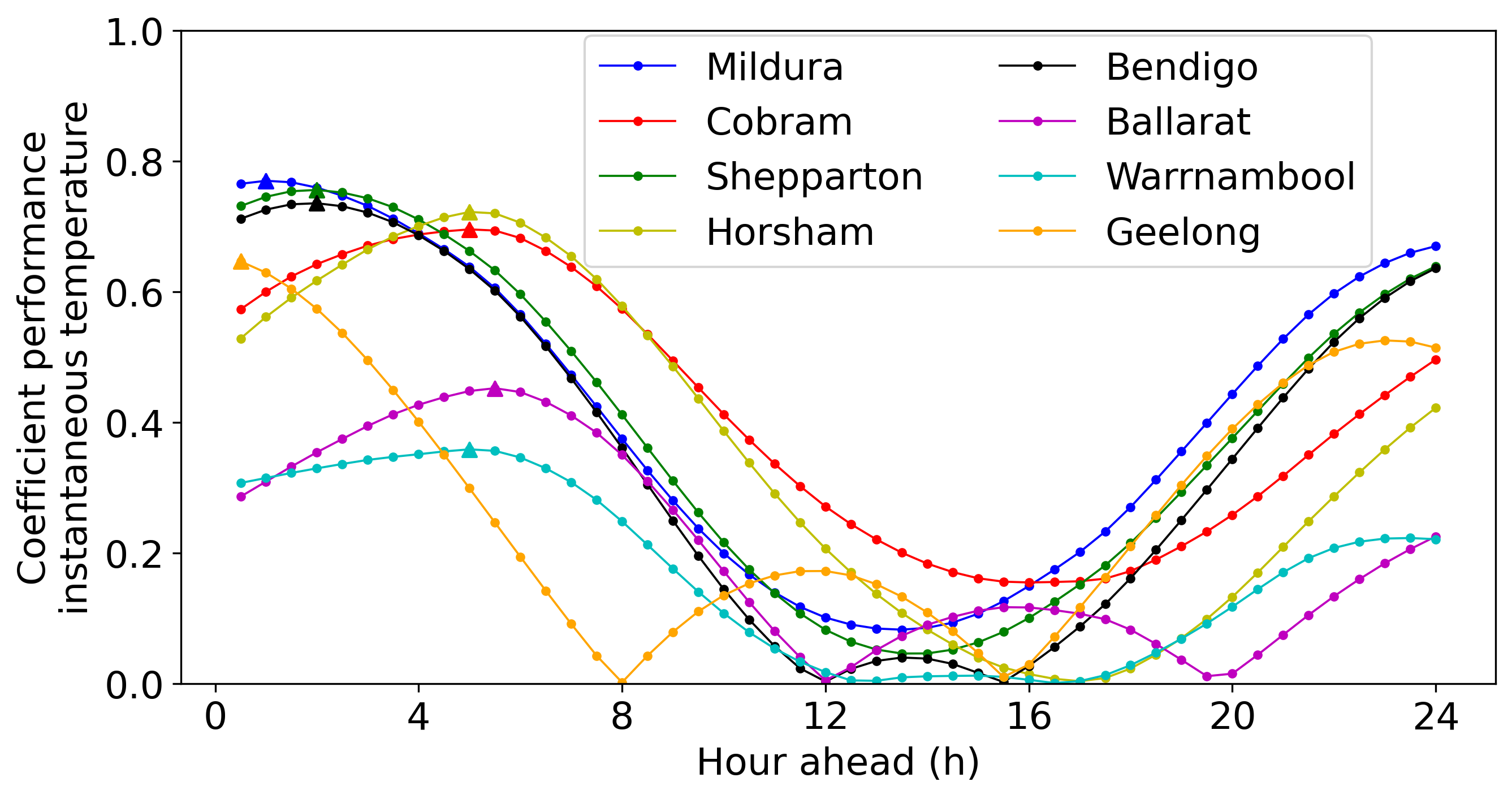}
\caption{Instantaneous temperature}
\label{fig:Step_4_2_a}
\end{subfigure}

\vspace{2ex}

\begin{subfigure}[b]{1.0\linewidth}
\includegraphics[width=\linewidth]{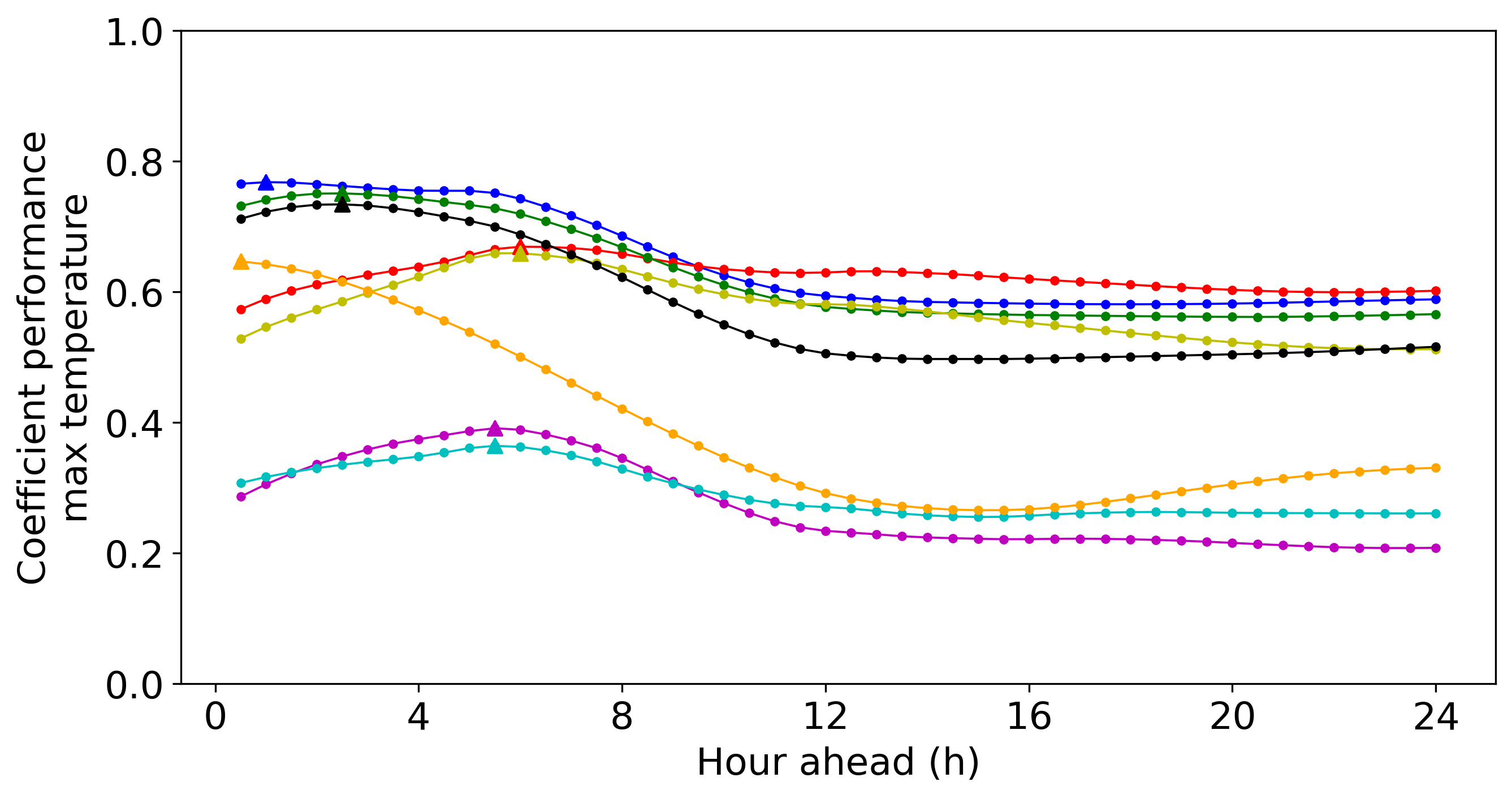}
\caption{Maximum temperature}
\label{fig:Step_4_2_b}
\end{subfigure}

\vspace{2ex}

\begin{subfigure}[b]{1.0\linewidth}
\includegraphics[width=\linewidth]{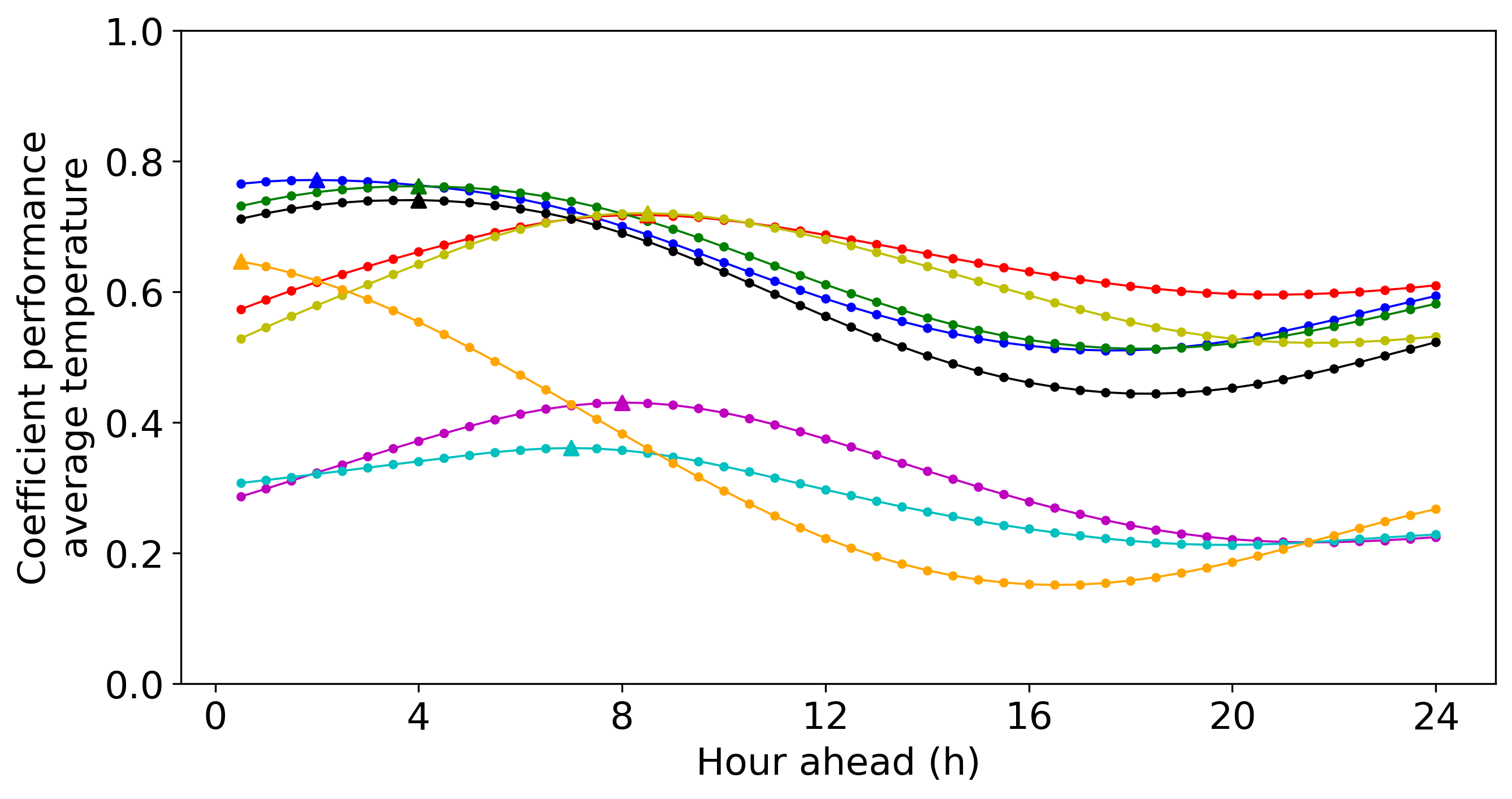}
\caption{Average temperature}
\label{fig:Step_4_2_c}
\end{subfigure}
\caption{Step 4 - load \& leading temperature correlation plots (0.5h interval) - 8 DNs in Victoria, Australia - wildfire seasons in 2015-2020}
\label{fig:Step_4_2}
\end{figure}

According to Step 3, the correlation analysis of energy consumption and temperature is first conducted to find the appropriate temperature conditions for each DN. The correlation analysis results for the instantaneous, maximum and average temperature at the eight DNs are displayed in Figure~\ref{fig:Step_4_2}. Though various DNs reach the strongest correlation coefficient at different hour-lead points, the eight places can be classified into two groups based on their correlation patterns. Mildura, Shepparton, Bendigo and Geelong are included in Group 1, where the strongest correlation coefficients for instantaneous, maximum and average temperature peak around 2-hour, 2-hour and 4-hour, respectively. For the other group: Cobram, Horsham, Ballarat and Warrnambool's correlation coefficients peak for instantaneous, maximum and average temperature conditions at around 4-hour, 4-hour and 8-hour, respectively. Regarding the two group patterns, 2-2-4-hour ahead and 4-4-8-hour ahead instantaneous, maximum and average temperature are defined as the three-temperature combination input condition, regarded as one condition to help forecast load using three temperatures. For the three-temperature combination condition, we want to test whether the different temperature condition combinations can improve the load forecast accuracy for data sets with different correlation patterns. For the identified temperature condition which produces the best correlation coefficient as found in Step 3, this condition is tested and applied to the other 7 DNs in Step 4. Therefore we have tested with no additional temperature condition, temperature condition with the best correlation coefficient and the three-temperature combination condition to promote model generalisation.

\begin{figure*}[htbp]
\centering
\includegraphics[width=1.95\columnwidth]{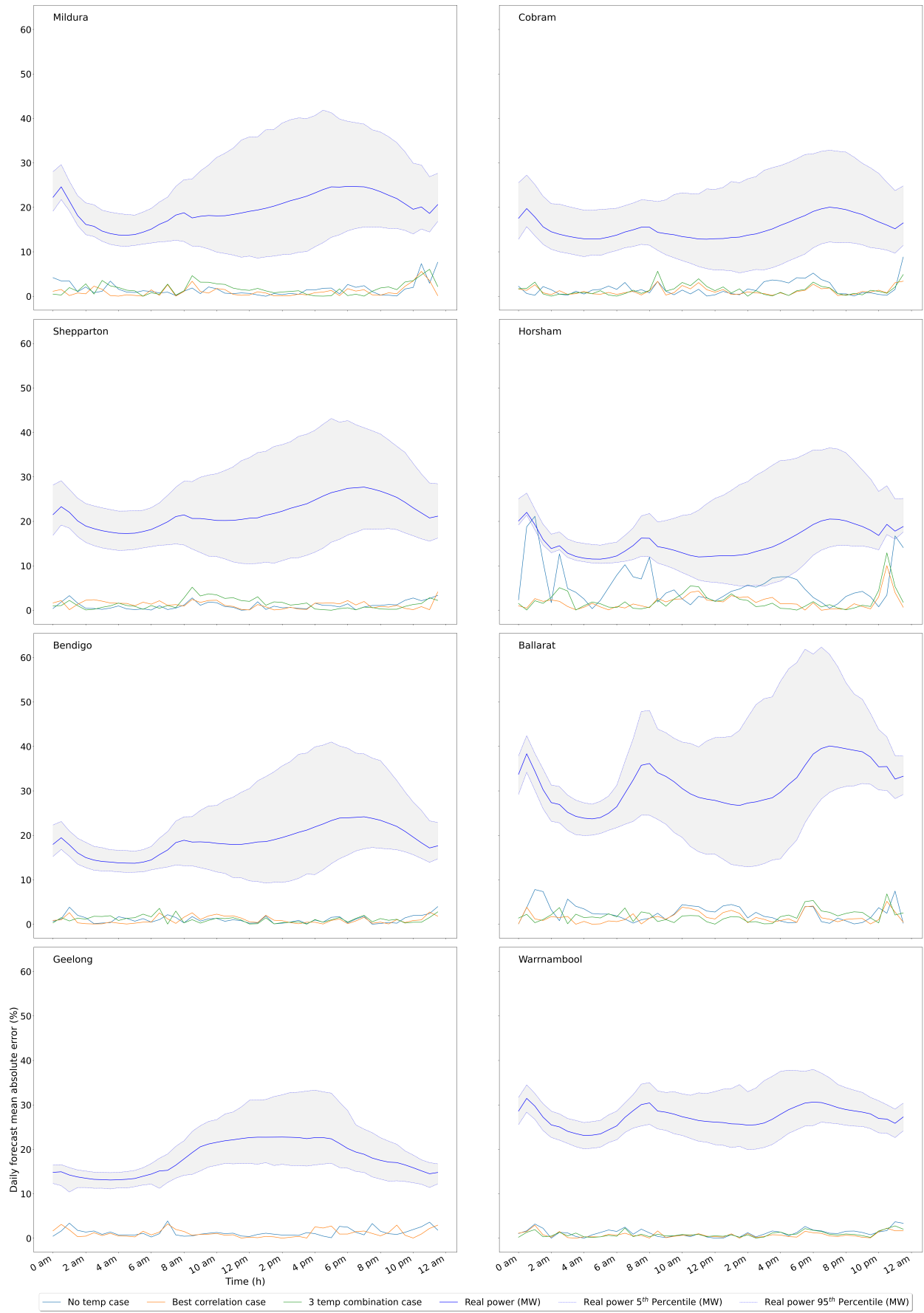}
\caption{Load forecast performance using 3 temperature conditions - daily mean absolute error - 8 DNs in Victoria, Australia - the wildfire seasons in 2015-2020}
\label{fig:Step_4_3}
\end{figure*}

As shown in Figure~\ref{fig:Step_4_3}, the real regional daily load curve with $5^{th}$ to $95^{th}$ percentile interval and the forecast performance of three temperature use cases are tested and compared among the 8 DNs. The daily average absolute error is plotted with the temporal resolution of 30 minutes, and the error variance, MSE and MAPE are assessed in Table~\ref{tbl2}. The four subfigures in the left column represent Group 1, whose strongest correlation pattern happened around 2-2-4 hours ahead. The four in the right column correlate to the peak around 4-4-8 hours ahead. There are two forecast lines for the Geelong case, while all other regions have three forecast lines. The reason is that the best instantaneous, maximum and average temperature correlation conditions are all 0.5 hours ahead in Geelong. Thus, the input of the best correlation coefficient temperature case is the same as the three-temperature combination, and we plot it as one situation.

\begin{table*}[width=2.06\linewidth,cols=10]
\caption{Performance metrics for 3 temperature conditions - 8 DNs}\label{tbl2}
\begin{tabular*}{\tblwidth}{@{} LLLLLLLLLL@{} }
\toprule
DNs & \multicolumn{3}{c}{$\sigma^{2}$ ($MW^{2}$)} & \multicolumn{3}{c}{MSE ($MW^{2}$)} & \multicolumn{3}{c}{MAPE (\%)}\\
\hline
Conditions & No temp & Best correl & 3 temps & No temp & Best correl & 3 temps & No temp & Best correl & 3 temps\\
\midrule
Mildura & 2.45 & 1.27 & 1.84 & 0.78 & 0.53 & 0.62 & 3.33 & 2.82 & 3.14 \\
Shepparton & 0.75 & 0.68 & 1.29 & 0.91 & 0.78 & 0.82 & 3.12 & 2.87 & 3.01 \\
Bendigo & 0.76 & 0.61 & 0.61 & 0.68 & 0.58 & 0.6 & 3.05 & 2.79 & 2.9 \\
Geelong & 0.77 & 0.79 & 0.79 & 0.97 & 0.94 & 0.94 & 3.63 & 3.52 & 3.52 \\
Cobram & 2.72 & 0.74 & 1.48 & 0.73 & 0.59 & 0.6 & 4.28 & 3.76 & 3.88 \\
Horsham  & 21.76 & 2.69 & 4.74 & 1.47 & 0.62 & 0.76 & 5.99 & 3.81 & 4.25 \\
Ballarat & 3.77 & 1.53 & 1.89 & 3.9 & 2.96 & 3.26 & 5.01 & 4.2 & 4.45 \\
Warrnambool & 0.74 & 0.38 & 0.42 & 0.82 & 0.66 & 0.68 & 2.49 & 2.16 & 2.23 \\
\bottomrule
\end{tabular*}
\end{table*}

\begin{figure*}[htbp]
\centering
\includegraphics[width=1.95\columnwidth]{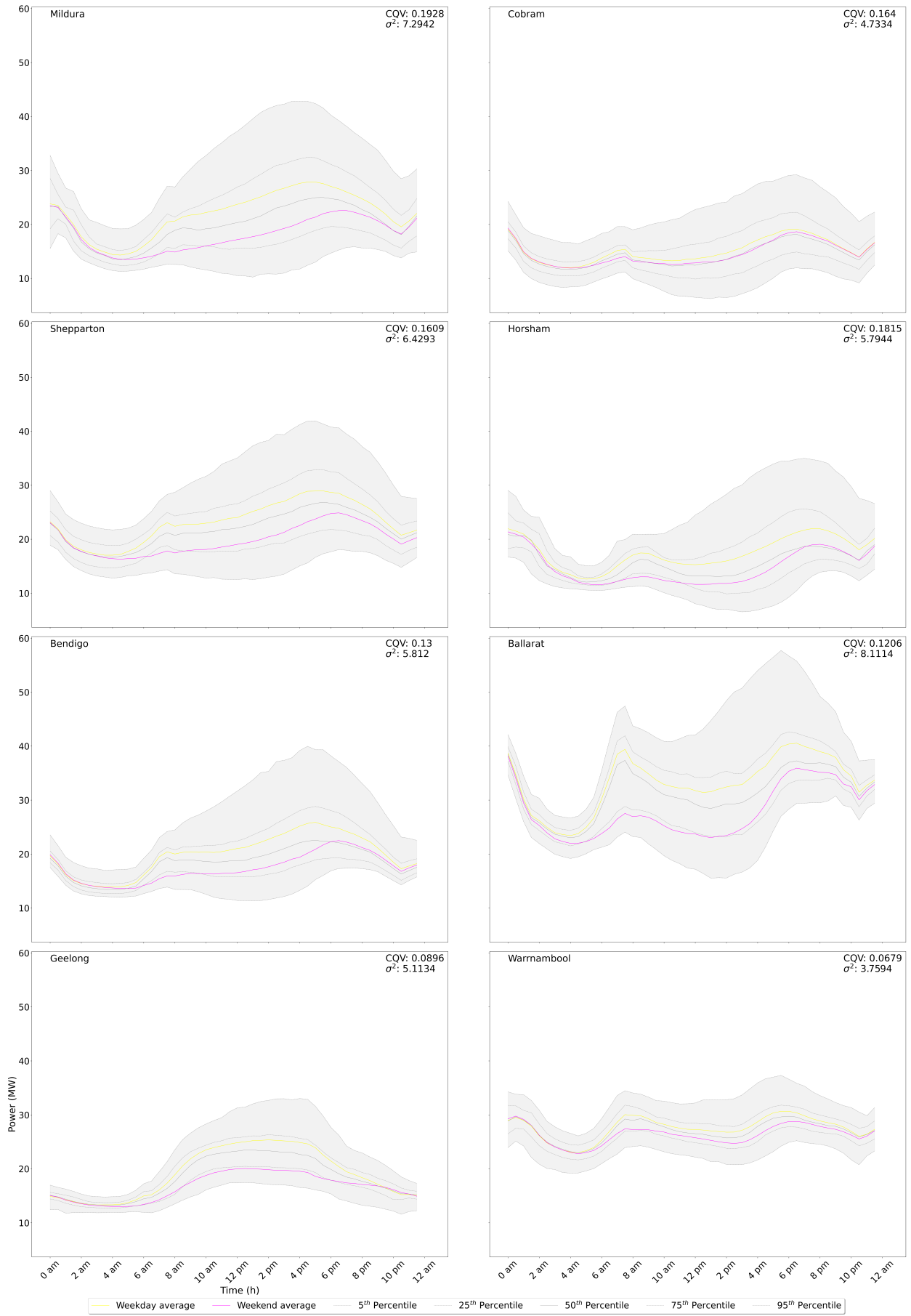}
\caption{Regional energy consumption dispersion analysis - 8 DNs in Victoria, Australia - the wildfire seasons in 2015-2020}
\label{fig:Step_4_4}
\end{figure*}

\begin{figure*}[htbp]
\centering
\includegraphics[width=1.95\columnwidth]{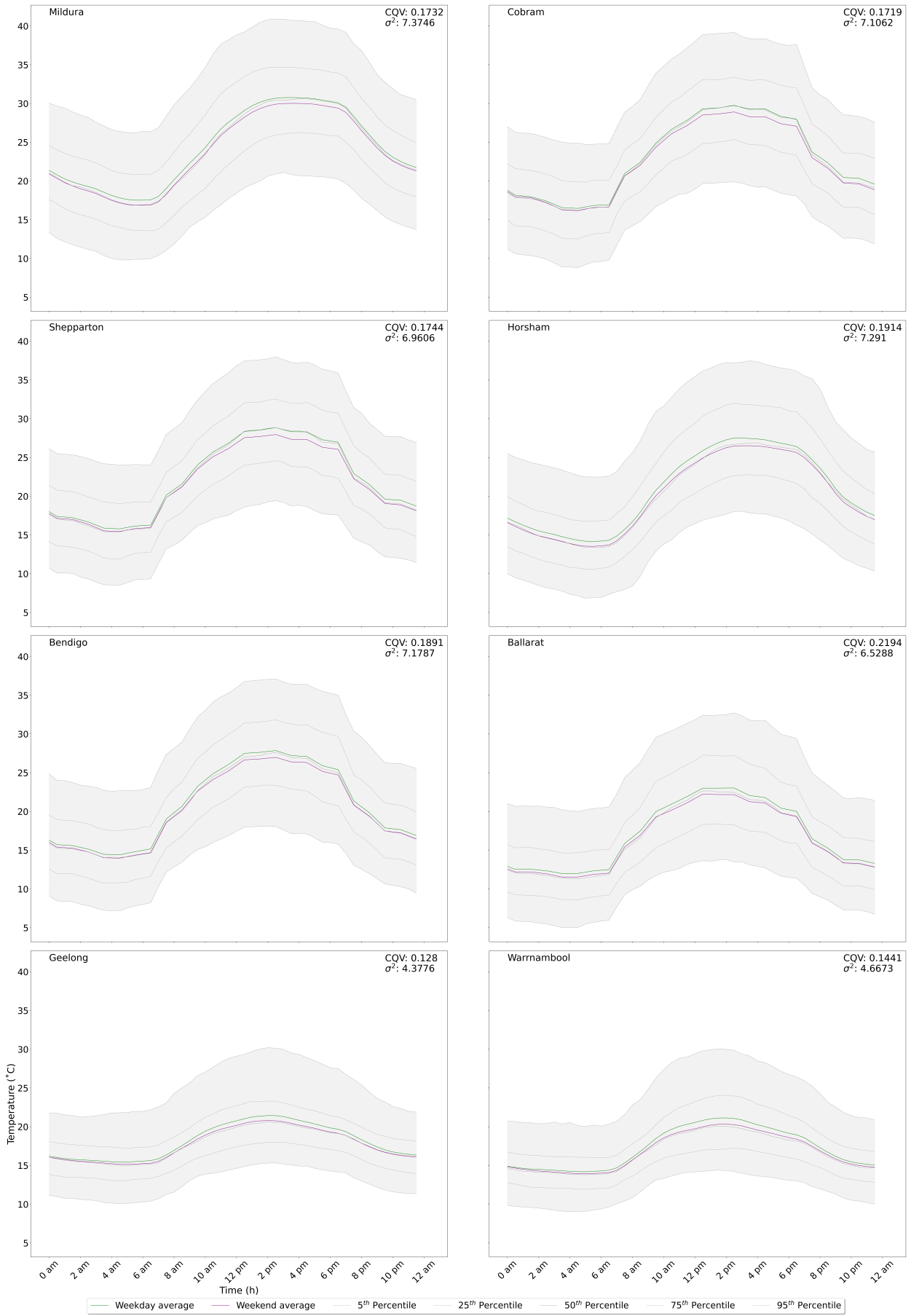}
\caption{Regional temperature dispersion analysis - 8 DNs in Victoria, Australia - the wildfire seasons in 2015-2020}
\label{fig:Step_4_5}
\end{figure*}

For the majority of 8 DNs, the best correlation coefficient temperature case outperforms the three-temperature combination case, and both exceed the no-temperature case. On average, the best correlation case performs 24.62\%, 7.26\% and 5.48\% better than the three-temperature combination case, and 42.90\%, 23.09\% and 14.11\% better than the no-temperature case for daily forecast error variance, MSE and MAPE, respectively. It is interesting that the forecast accuracy improvements by applying temperature to the model for Group 2 DNs are more evident in all three error metrics. This indicates that the temperature plays a more important role in improving load forecast accuracy in locations in Group 2. Thus, we need to further analyse the feature differences between Group 1 and Group 2 data sets to find why the forecast performance has different sensitivities to the participation of temperature in different DN group cases.\\
As temperature and energy load data are the two main inputs for our generalised forecast model, the wildfire season temperature and energy data sets are first visualised and assessed to explore the data features of the two groups. In Figure~\ref{fig:Step_4_4} and Figure~\ref{fig:Step_4_5}, the weekday average, weekend average and daily percentile curves are displayed to show the regional load and regional temperature data set dispersion level, respectively. To quantify the variability of the regional load and temperature, the CQV and ${\sigma}^{2}$ (error variance) are calculated for each DN. To be consistent with the forecast performance comparison plot in Figure~\ref{fig:Step_4_3}, the four subfigures in the left column describe the data set dispersion level for Group 1 DNs, and the four on the right represent Group 2 DNs' analysis in both Figure~\ref{fig:Step_4_4} and Figure~\ref{fig:Step_4_5}. For energy data sets, the average ${\sigma}^{2}$ and CQV of Group 1 DNs are 9.12\% and 6.84\% higher than in Group 2. And for temperature dispersion level, the average ${\sigma}^{2}$ of Group 1 is 1.15\% higher than Group 2, while the CQV of Group 2 is 9.34\% higher than Group 1. It is concluded that the load performance improvement potential by temperature is mainly affected by the degree of dispersion level of load data sets: the more densely distributed the regional load data, the better the help from temperature participating in the load forecast.


\begin{figure}
	\centering
            \includegraphics[width=.9\linewidth]{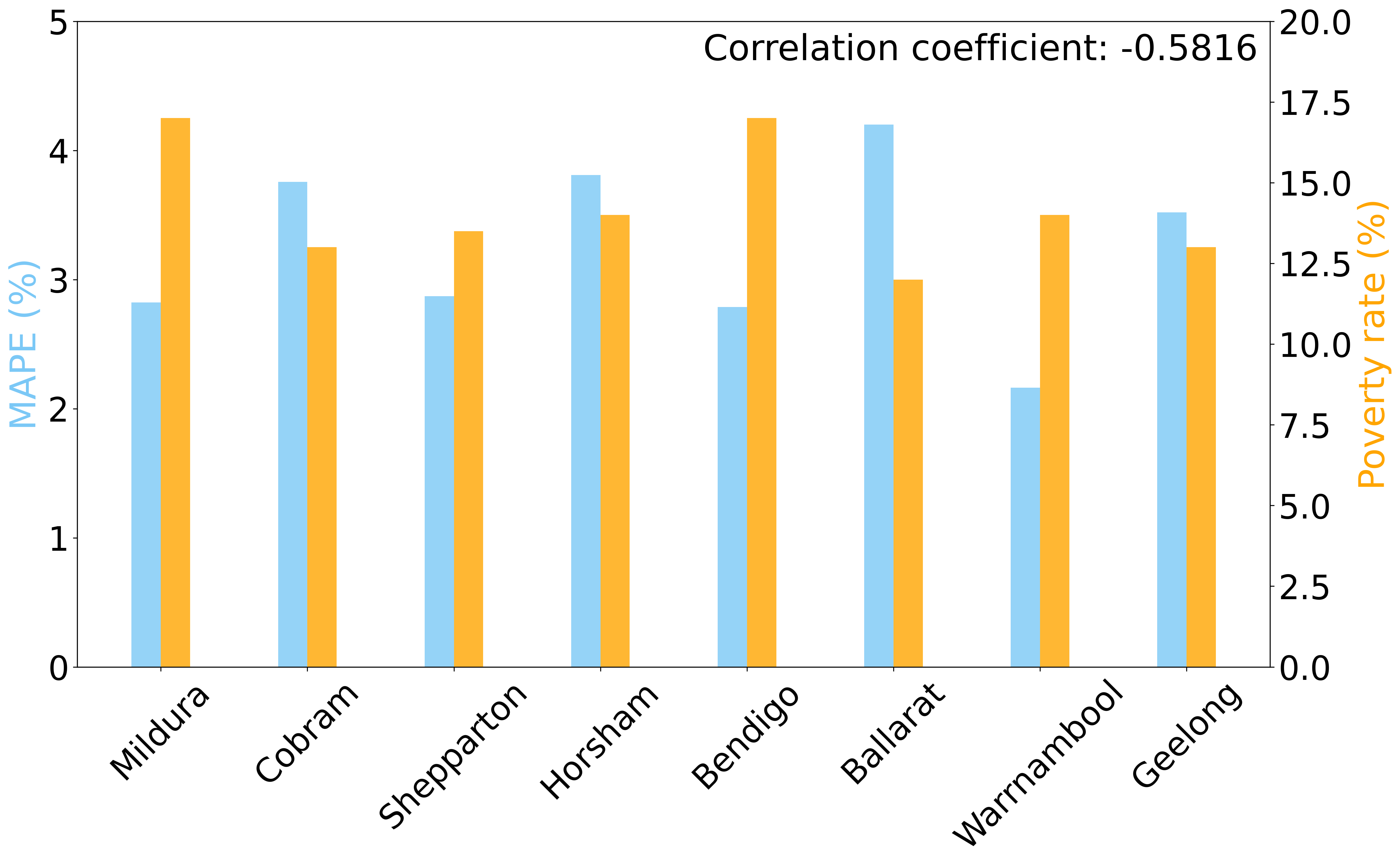}
	\caption{Load forecast MAPE and poverty rate at 8 DNs in Victoria, Australia - the wildfire seasons in 2015-2020}
	\label{fig:Step_4_6}
\end{figure}

In addition, more regional features are studied to discuss potential factors affecting our model generalisation performance. Factors like GDP, GDP per capita, population density and poverty rate are tested for 8 DNs, to find the potential relationship between factors and the forecast accuracy (MSE \& MAPE). Among all these regional features, our results shed new light on the inverse relationship between the MAPE and poverty rate. In Figure~\ref{fig:Step_4_6}, the regional MAPE and poverty rate for 8 DNs are displayed in a bar chart, showing the negative relationship with the correlation coefficient of -0.5816. The relationship indicates that the poorer the region, the simper the energy consumption behaviour with less forecast error. In contrast, richer residences are not restricted to the peak/off-peak tariff, turning on and off switches more randomly and thus harder to predict. In real operation, the negative correlation between forecast error and poverty rate can be considered when doing the resource pre-dispatch and long-term planning, i.e., leaving more space when forecasting energy demand in a higher-income area.

\subsection{Robustness analysis and other uncertain factor discussion}
We have chosen 2015-2020 as the main research period. The Australian annual wildfire season runs from October to March. The first three-year wildfire seasons: 15-16, 16-17 and 17-18 are used as the training data sets for our forecast model. The forecast performance is then evaluated using the last two-year wildfire season data. In order to validate the robustness of data selection, some other methods of data set selection are tested: training and testing data periods are randomly rotated to compare the forecast accuracy stability. As discussed in Step 4, 8 DNs are classified into two groups. The forecast performance improvement of Group 2 is more sensitive to temperature participation than Group 1. Thus, one DN from each group is tested: Shepparton from Group 1 and Warrnambool from Group 2. The random selections of training and testing data sets and their corresponding MAPE are displayed in Table~\ref{tbl3}. For both DNs, the forecast performs relatively better when not using 2015-2016 as the training data set. The highest two MAPEs for each DN appear in the $1^{st}$ and the $5^{th}$ rows, indicating that using 2015-2016 as a training data set weakens the model forecast performance. As 2015-2016 was an El Niño impact year \cite{ECMWF2017}, it may be harder to capture the relationship between load and temperature using these periods data as training input. In the real operation, we consider it better not to use El Niño or La Niña impact years as training data sets, especially when climate factors are considered in the model. Regarding the magnitude, the MAPEs of the four test cases are still within an acceptable range of the original case. El Niño or La Niña impact years should be avoided as the training data sets to enhance the robustness of the model training. Overall, the random selection of training and testing data sets does not affect the performance significantly though, proving our model's robustness to encountered large-scale climate variability.

\begin{table*}[width=2.06\linewidth,cols=8]
\caption{Random selection of training and testing data sets - Shepparton \& Warrnambool}\label{tbl3}
\begin{tabular*}{\tblwidth}{@{} LLLLLLLL@{} }
\toprule
Year & 15-16 & 16-17 & 17-18 & 18-19 & 19-20 & Shepparton MAPE (\%) & Warrnmabool MAPE (\%)\\
\midrule
Original & Train & Train & Train & Test & Test & 2.87 & 2.16 \\
Test 1 & Test & Test & Train & Train & Train & 1.96 & 1.85 \\
Test 2 & Test & Train & Test & Train & Train & 2.19 & 1.68 \\
Test 3 & Test & Train & Train & Test & Train & 2.12 & 2.01 \\
Test 4 & Train & Test & Train & Train & Test & 2.56 & 2.07 \\
\bottomrule
\end{tabular*}
\end{table*}

As demonstrated in Figure~\ref{fig:Step_4_3}, the midnight energy consumption uptick can be found in almost all DNs in Victoria. It is caused by switching on of off-peak electrical hot water systems \cite{SustainabilityVic2022}. It is also noticeable that the average error during the midnight load uptick period is relatively higher compared with the rest of the day for many DNs. This indicates that, the forecast performance excluding the midnight period is worth studying to explore the potential performance improvement methods. According to Figure~\ref{fig:Step_4_3}, the average absolute error in Horsham peaks around midnight, while the error lines are relatively stable all day in Warrnambool. Thus, Warrnambool and Horsham are picked to test the performance change with and without considering the off-peak electrical hot water system period. The off-peak electrical hot water switching-on period is fixed, but the scale is human-controlled and changeable. After taking out the midnight period, the best correlation temperature condition MAPE drops from 2.16\% to 2.10\% in Warrnambool and decreases from 3.81\% to 3.70\% in Horsham. On average, the forecast error is reduced by 2.84\%, which indicates that the model accuracy can be improved by knowing the off-peak electrical hot water use plan in advance.

\subsection{Cost-benefit analysis} \label{Cost-bene}
As mentioned in Step 4, the load forecast performance is improved by different percentages in daily error variance, MSE and MAPE after applying our flexible correlation-based temperature conditions. As MAPE has the unit of "\%", the MAPE decrement after applying our method is used to estimate the cost-benefit.\\
According to Figure~\ref{fig:Step_4_3}, the average MAPE using the no-temperature forecast is 3.86\% for 8 DNs during the wildfire seasons in 2015-2020, whereas the average MAPE using the three-temperature combination condition is 3.42\%, and the average MAPE of the best correlation coefficient temperature condition is 3.24\%. The forecast error is reduced by 11.40\% and 16.06\% when using the three-temperature combination method and the best correlation temperature method. According to the Australian Energy Regulator \cite{AER}, the annual electricity consumption of Victoria in the 2019-2020 financial year is 44.3TWh. The load forecast error may be reduced by 0.19TWh and 0.27TWh using the three-temperature combination method and the best correlation coefficient temperature method, respectively. Australia's national average residential electricity tariff is 29.8c/KWh from 2019-2020 \cite{AEMC}. The total cost saving possible for Victoria is calculated by multiplying the forecast error decrement by the electricity tariff. This gives indicated annual total savings of AU\$56.62 million and AU\$80.46 million by using the three-temperature combination method and the best correlation coefficient temperature method, respectively.

\section{Conclusion and future work}
Our paper proposes a systematic and robust multi-factor GRU-based load forecast model to predict short-term DN electricity consumption in Victoria, Australia, during the extreme wildfire seasons in 2015-2020. Factors affecting load forecast, i.e., RNN input data lengths, calendar labels, and preceding temperatures, have been further analysed and exploited to maximise their functions in advancing load forecast accuracy through three main steps. After testing our model on 8 DNs in Victoria, Australia, our model is proved to be able to achieve good and stable performance during extensive climate variability periods, with a forecast error of around 3\% MAPE for DNs in Victoria State.\\
In Step 1, only historical load is input into the RNN model. The input data length is adjusted to find the appropriate input structure with both good accuracy and short training time. MSE and MAPE drop to a stable level when considering the current to pre-16-hour information as the model input.\\
In Step 2, different calendar label types are attached to the model input. No-day-type-label, three-day-type-label and eight-day-type-label methods are tested with the Horsham DN load. The three-day-type-label method classifies days into weekdays, weekends and holidays. In contrast, the eight-day-type-label method represents days from Monday to Sunday each, plus a holiday label. Regarding MAPE, the eight-day-type-label method stands out among all three cases, 4.23\% and 1.01\% better than the no-day-type-label and three-day-type-label methods, respectively.\\
A flexible temperature condition method is proposed in Step 3. Firstly, the relationship between load and different types of leading temperatures have been investigated. The pre-n-hour instantaneous temperature, the maximum temperature in pre-n-hour and the average temperature in pre-n-hour are assessed based on the Pearson Correlation Coefficient and the $2_{nd}$ order $r^{2}$ value. Six temperature conditions with the highest Pearson Correlation Coefficients or the highest $2_{nd}$ order $r^{2}$ values and the simultaneous-temperature-load case plus the no-temperature case are input into the GRU forecast model to find the temperature condition that best help with energy load forecast. For the Horsham data sets, the best correlation temperature condition case with GRU structure performs the best among all cases, decreasing the MAPE by 30.73\% and 36.39\% compared to the simultaneous-temperature-load method and the no-temperature method, respectively.\\
Our forecast model is further developed and generalised by testing it in the real DNs in Victoria in Step 4. According to patterns in the load-temperature correlation analysis plots, 8 DNs are classified into two groups. The best correlation temperature conditions, three-temperature combination conditions and no-temperature conditions are input into the 8 DN forecast models. The best correlation case performs better than using the three-temperature combination condition and then better than the no-temperature condition in the daily average error variation, MSE and MAPE. The dispersion level of temperature and energy data sets are assessed and visualised to understand further the inner reason affecting forecast performances. It is noticeable that the more densely distributed regional load data, the more important flexible temperature conditions participating in load forecast.\\
Cost-benefit and model robustness analyses are assessed to demonstrate the significant economic benefit, and model stability to large-scale climate variability. It is shown that AU\$56.62 million and AU\$80.46 million may be saved by applying the three-temperature combination condition method and the best correlation coefficient leading temperature condition as the input, respectively. In the robustness analysis, the forecast results stay stable when randomly rotating the training and testing data sets, keeping the same magnitude of errors.\\
To further develop the load forecast model, some other factors are discovered that possibly contribute to better forecast accuracy in the future. MAPE and regional poverty rate have a relatively strong negative correlation (-0.5816), which should be considered in the regional load forecast in real operations. The model robustness analysis demonstrates that El Niño or La Niña impact years should be avoided as training data sets to guarantee the load forecast performance's stability. According to the regional load dispersion level plots, the midnight energy use upticks for most DNs in Victoria are caused by switching on of the off-peak electrical hot water systems. The operational period of the off-peak electrical use is fixed, but the scale is manually set and different from day to day. Hence, the performance with and without the on-period is computed and compared, proving that the forecast error can be decreased by 2.84\% by knowing the energy use plan in advance. Overall, the regional poverty rate, the climate impact year and the manually-controlled large-scale off-peak energy use plan should be considered in the load forecast to improve forecast performance in the future.\\
In conclusion, our paper proposes the novel GRU-based load forecast model considering varying input data structures, calendar effects and flexible leading temperature conditions to forecast DN load in Victoria, Australia, during the wildfire seasons in 2015-2020. Our paper maximises multi-factor functions to enhance the load forecast performance, showing good forecast accuracy compared to other short-term DN load forecast models. With the climate robustness and economic benefits being well-proved, the standardised load forecast method developed in our paper can be applied to strengthen DNs suffering from other types of extreme weather in other regions.

\bibliographystyle{model1-num-names}

\bibliography{cas-refs}

\printcredits

 \clearpage

\section{Appendix}
\label{Appen}

\end{document}